\RequirePackage{fix-cm} 
\documentclass[a4paper, twoside, reqno, dvips, 12pt]{amsart}
\usepackage{fixltx2e}   

\usepackage[latin1]{inputenc}
\usepackage[T1]{fontenc}

\usepackage{eucal}
\usepackage{esint}
\usepackage{xspace}
\usepackage{amsgen}
\usepackage{amssymb}
\usepackage{amsmath}
\usepackage{amsfonts}
\usepackage{MnSymbol}
\usepackage[nice]{nicefrac}
\usepackage{mathrsfs, dsfont}

\usepackage{a4wide}

\usepackage[table]{xcolor}

\definecolor{oneblue}{rgb}{0.0, 0.0, 0.85}
\definecolor{darkgrey}{rgb}{0.273, 0.281, 0.30}
\definecolor{Lightgray}{rgb}{0.89, 0.89, 0.89}
\definecolor{Lightblue}{RGB}{214, 214, 214}
\definecolor{bckg}{RGB}{20.8, 20.8, 20.8} 

\usepackage{psfrag}
\usepackage{graphicx}
\usepackage{subfigure}
\usepackage{morefloats}
\usepackage{indentfirst}

\usepackage{acronym}
\usepackage{microtype}
\usepackage[labelfont={bf,sf},%
            labelsep=period,%
            justification=raggedright]{caption}

\usepackage[usenames, dvipsnames, pdf]{pstricks}
\usepackage{epsfig}
\usepackage{pst-grad} 
\usepackage{pst-plot} 

\usepackage[colorlinks,
            urlcolor=oneblue,
            linkcolor=oneblue,
            citecolor=oneblue,
            bookmarksopen=false,
            pagebackref]{hyperref}

\usepackage[explicit]{titlesec}

\titleformat{\section}
  {\color{darkgrey}\Large\sffamily\bfseries}
  {}
  {0em}
  {\colorbox{bckg!5}{\parbox{\dimexpr\linewidth-2\fboxsep\relax}{\centering\thesection. #1}}}
  []

\titleformat{name=\section,numberless}
  {\color{darkgrey}\Large\sffamily\bfseries}
  {}
  {0em}
  {\colorbox{bckg!10}{\parbox{\dimexpr\linewidth-2\fboxsep\relax}{\centering#1}}}
  []

\titleformat{\subsection}
  {\color{darkgrey}\large\sffamily\bfseries}
  {}
  {0em}
  {\colorbox{bckg!5}{\parbox{\dimexpr\linewidth-2\fboxsep\relax}{\centering\thesubsection. #1}}}
  []

\titleformat{name=\subsection,numberless}
  {\color{darkgrey}\Large\sffamily\bfseries}
  {}
  {0em}
  {\colorbox{bckg!10}{\parbox{\dimexpr\linewidth-2\fboxsep\relax}{\centering#1}}}
  []

\titleformat{\subsubsection}
  {\sffamily\normalsize\bfseries}
  {\thesubsubsection}
  {0em}
  {#1}
  []

\titleformat{\paragraph}[runin]
  {\sffamily\small\bfseries}
  {}
  {0em}
  {#1}

\titlespacing*{\section}{1.0em}{1.0em}{0.8em}[0em]
\titlespacing*{\subsection}{1.0em}{1.0em}{0.8em}[0em]
\titlespacing*{\subsubsection}{1.0em}{0.7em}{0.6em}[0em]

\usepackage{titletoc}

\contentsmargin{0.55em}

\newlength{\tocsep}
\titlecontents{section}[\tocsep]
  {\addvspace{10pt}\bfseries\sffamily}
  {\contentslabel[\thecontentslabel]{\tocsep}}
  {}
  {\ \titlerule*[0.75pc]{.}\ \thecontentspage}
  []
\titlecontents{subsection}[\tocsep]
  {\addvspace{8pt}\sffamily}
  {\contentslabel[\thecontentslabel]{\tocsep}}
  {}
  {\ \titlerule*[0.5pc]{.}\ \thecontentspage}
  []
\titlecontents{subsubsection}[\tocsep]
  {\addvspace{7pt}\footnotesize\sffamily}
  {\contentslabel[\thecontentslabel]{\tocsep}}
  {}
  {\ \titlerule*[0.5pc]{.}\ \thecontentspage}
  []

\usepackage{fancyhdr}
\usepackage{lastpage}

\newcommand*\Title{Numerical study of gKG equations}
\newcommand*\Authors{D.~Dutykh, M.~Chhay \& D.~Clamond}
\newcommand*{\plogo}{{\texttt{arXiv.org} / \textsc{hal}}} 

\usepackage{acronym}
\acrodef{cz}[cZ]{cubic Zakharov}
\acrodef{ivp}[IVP]{Initial Value Problem}
\acrodef{gkg}[gKG]{generalized Klein--Gordon}
\acrodef{NLS}[NLS]{Nonlinear Schr\"{o}dinger Equation}

\numberwithin{equation}{section}

\newtheorem{remark}{Remark}

\newcommand{\M}{\mathbb{M}}
\newcommand{\K}{\mathbb{K}}
\newcommand{\R}{\mathds{R}}
\newcommand{\J}{\mathbb{J}}
\newcommand{\dy}{\partial_y}
\newcommand{\dt}{\partial_t}

\newcommand{\F}{\mathcal{F}}

\newcommand{\Ll}{\mathcal{L}}

\renewcommand{\O}{\mathcal{O}}
\renewcommand{\H}{\mathcal{H}}
\renewcommand{\S}{\mathcal{S}}
\newcommand{\eps}{\varepsilon}

\newcommand{\x}{\boldsymbol{x}}

\newcommand{\sech}{\mathrm{sech}}
\renewcommand{\k}{\boldsymbol{k}}
\renewcommand{\u}{\boldsymbol{u}}
\newcommand{\uu}{V}

\newcommand{\vv}{W}
\newcommand{\grad}{\boldsymbol{\nabla}}

\newcommand{\s}{\mathsf{s}}

\newcommand{\half}{{\textstyle{1\over2}}}

\newcommand{\quat}{{\textstyle{1\over4}}}

\newcommand{\ud}{\mathrm{d}}

\newcommand{\ue}{\mathrm{e}}
\newcommand{\ui}{\mathrm{i}}

\newcommand{\vF}{\boldsymbol{F}} 
\newcommand{\vu}{\boldsymbol{u}}                   
\newcommand{\vx}{\boldsymbol{x}}                               
\newcommand{\nab}{\boldsymbol{\nabla}}            
\newcommand{\scal}{\boldsymbol{\cdot}}
\newcommand{\valpha}{\boldsymbol{\alpha}}
\newcommand{\vgamma}{\boldsymbol{\gamma}}
\newcommand{\vmu}{\boldsymbol{\mu}}     

\newcommand{\vtau}{\boldsymbol{\tau}}  

\newcommand{\us}{\tilde{u}}
\newcommand{\vs}{\tilde{v}}

\newcommand{\nus}{\tilde{\nu}}
\newcommand{\phis}{\tilde{\phi}}
\newcommand{\vus}{\tilde{\boldsymbol u}}
\newcommand{\vmus}{\tilde{\boldsymbol \mu}}

\renewcommand{\Re}{\operatorname{Re}}

\newcommand{\ie}{\emph{i.e.}~}

\newcommand{\etal}{\emph{et al.}~}

\begin{document}

\title[\Title]{Numerical study of the generalised Klein--Gordon equations}

\author[D.~Dutykh]{Denys Dutykh$^*$}
\address{LAMA, UMR 5127 CNRS, Universit\'e Savoie Mont Blanc, Campus Scientifique, 73376 Le Bourget-du-Lac Cedex, France}
\thanks{$^*$ Corresponding author}
\email{Denys.Dutykh@univ-savoie.fr}
\urladdr{http://www.denys-dutykh.com/}

\author[M.~Chhay]{Marx Chhay}
\address{LOCIE, UMR 5271 CNRS, Universit\'e Savoie Mont Blanc, Campus Scientifique, 73376 Le Bourget-du-Lac Cedex, France}
\email{Marx.Chhay@univ-savoie.fr}
\urladdr{http://marx.chhay.free.fr/}

\author[D.~Clamond]{Didier Clamond}
\address{Universit\'e de Nice -- Sophia Antipolis, Laboratoire J. A. Dieudonn\'e, Parc Valrose, 06108 Nice cedex 2, France}
\email{diderc@unice.fr}
\urladdr{http://math.unice.fr/~didierc/}


\begin{titlepage}
\setcounter{page}{1}
\thispagestyle{empty} 
\noindent
{\Large Denys \textsc{Dutykh}}\\
{\it\textcolor{gray}{CNRS--LAMA, University Savoie Mont Blanc, France}}\\[0.02\textheight]
{\Large Marx \textsc{Chhay}}\\
{\it\textcolor{gray}{LOCIE, University Savoie Mont Blanc, France}}\\[0.02\textheight]
{\Large Didier \textsc{Clamond}}\\
{\it\textcolor{gray}{University of Nice Sophia Antipolis, France}}\\[0.16\textheight]

\colorbox{Lightblue}{
  \parbox[t]{1.0\textwidth}{
    \centering\huge\sc
    \vspace*{0.7cm}

    Numerical study of the generalised Klein-Gordon equations

    \vspace*{0.7cm}
  }
}

\vfill 

\raggedleft     
{\large \plogo} 
\end{titlepage}


\newpage
\maketitle
\thispagestyle{empty}

\begin{abstract}
In this study, we discuss an approximate set of equations describing water wave propagating in deep water. These \acf{gkg} equations possess a variational formulation, as well as a canonical Hamiltonian and multi-symplectic structures. Periodic travelling wave solutions are constructed numerically to high accuracy and compared to a seventh-order Stokes expansion of the full Euler equations. Then, we propose an efficient pseudo-spectral discretisation, which allows to assess the stability of travelling waves and localised wave packets.

\bigskip
\noindent \textbf{\keywordsname:} deep water approximation; spectral methods; travelling waves; periodic waves; stability
\end{abstract}

\maketitle


\newpage
\tableofcontents
\thispagestyle{empty}
\newpage


\section{Introduction}

The water wave problem counts today more than 200 years of history (see A.~\textsc{Craik} (2004), \cite{Craik2004}). Despite some recent progress \cite{Dias2006a, Dyachenko1996, Fructus2005, Grilli}, the complete formulation remains a mathematical difficult problem and a stiff numerical one. Consequently, researchers have always been looking for specific physical regimes which would allow to simplify the governing equations \cite{Stoker1958, Whitham1999}. There are two main regimes which attracted a particular attention from the research community: shallow and deep water approximations \cite{Johnson1997, Mei1994}.

If $\lambda$ is a characteristic wavelength and $h$ is an average water depth, the shallow water approximation consists to assume that $h/\lambda \ll 1$ or in other words, the water depth is much smaller compared to the typical wavelength. This regime is relevant in coastal engineering problems \cite{Ma2010, Sorensen1997, Wu1981}. In open ocean only tsunami and tidal waves are in this regime \cite{Dias2006, Kervella2007}.

The deep water approximation is exactly the opposite case when $h/\lambda \gg 1$, \ie the water depth is much bigger than the typical wavelength. In practice, some deep water effects (defocussing type of the NLS equation) can already manifest when $k h = 2\pi{h}/{\lambda} \gtrsim 1.36$. This regime is relevant for most wave evolution problems in open oceans \cite{Boccotti2000}. In the present paper, we present a detailed derivation of what we call a ``\acf{gkg}'' equations using a variational principle \cite{Clamond2009}. To our knowledge, it is a novel model in deep water regime. By making comparisons with the full Euler equations, we show that these equations can, on some peculiar features, outperform the celebrated \acf{cz} equations \cite{Korotkevich2007, Korotkevich2008}. Recently, a novel so-called \emph{compact Dyachenko--Zakharov} equation was proposed \cite{Dyachenko2011} which describes the evolution of the complex wave envelope amplitude in deep waters. This promising equation results from a sequence of thoroughly chosen canonical transformations, making the direct comparisons rather tricky.

The \acs{gkg} equations have multiple variational structures. First of all, they appear as Euler--Lagrange equations of an approximate Lagrangian that possesses also a canonical Hamiltonian formulation \cite{Clamond2009}. In this study, we show that the \acs{gkg} system can be recast into the multi-symplectic form \cite{Bridges2001, Marsden1998} as well. The main idea behind this formulation is to treat the time and space variables on equal footing \cite{Bridges2006} while, for instance in Hamiltonian systems, the time variable is privileged with respect to the space. Based on this special structure, numerous multi-symplectic schemes have been proposed for multi-symplectic PDEs including the celebrated KdV and NLS equations \cite{Bridges2001, Moore2003, Schober2008, Zhao2000}. These schemes are specifically designed to preserve exactly the discrete multi-symplectic form. However, these schemes turn out to be fully implicit, thence advantageous only for long time simulations using large time steps. Since in the present study we focus on the mid-range dynamics, we opt for a pseudo-spectral method which can insure a high accuracy with an explicit time discretisation \cite{Clamond2001, Fructus2005, Milewski1999, Trefethen2000}. Since the periodic and localised solutions play an important role in the nonlinear wave dynamics \cite{Osborne2010}, we use the numerical method to study the behaviour of these solutions.

The present paper is organised as follows. In Section~\ref{sec:model} we briefly present the essence of the deep water approximation and derive the \acs{gkg} equations. In Secton~\ref{sec:struct}, we discuss some structural properties of the model and, in Section~\ref{sec:compa}, we compare it to the classical \acf{cz} equations. Periodic travelling wave solutions are computed in Section~\ref{sec:tw}. The numerical method for the \acs{gkg} initial value problem is described in Section~\ref{sec:nums}. Some numerical tests are presented in Section~\ref{sec:numres}. Finally, the last Section~\ref{sec:concl} contains main conclusions of this study.

\section{Mathematical modelling}\label{sec:model}

\begin{figure}
\begin{center}
\scalebox{0.9} 
{
\begin{pspicture}(0,-2.83)(15.819062,2.83)
\definecolor{color1}{rgb}{0.03529411764705882,0.1568627450980392,0.9490196078431372}
\pscustom[linewidth=0.06,linecolor=color1]
{
\newpath
\moveto(0.1,0.67)
\lineto(0.37,0.79)
\curveto(0.505,0.85)(0.73,0.945)(0.82,0.98)
\curveto(0.91,1.015)(1.07,1.035)(1.14,1.02)
\curveto(1.21,1.005)(1.35,0.955)(1.42,0.92)
\curveto(1.49,0.885)(1.645,0.785)(1.73,0.72)
\curveto(1.815,0.655)(2.02,0.535)(2.14,0.48)
\curveto(2.26,0.425)(2.505,0.39)(2.63,0.41)
\curveto(2.755,0.43)(3.005,0.515)(3.13,0.58)
\curveto(3.255,0.645)(3.455,0.79)(3.53,0.87)
\curveto(3.605,0.95)(3.775,1.07)(3.87,1.11)
\curveto(3.965,1.15)(4.215,1.155)(4.37,1.12)
\curveto(4.525,1.085)(4.755,1.0)(4.83,0.95)
\curveto(4.905,0.9)(5.1,0.745)(5.22,0.64)
\curveto(5.34,0.535)(5.565,0.355)(5.67,0.28)
\curveto(5.775,0.205)(6.015,0.18)(6.15,0.23)
\curveto(6.285,0.28)(6.555,0.445)(6.69,0.56)
\curveto(6.825,0.675)(7.04,0.84)(7.12,0.89)
\curveto(7.2,0.94)(7.395,0.99)(7.51,0.99)
\curveto(7.625,0.99)(7.88,0.905)(8.02,0.82)
\curveto(8.16,0.735)(8.43,0.585)(8.56,0.52)
\curveto(8.69,0.455)(8.935,0.405)(9.05,0.42)
\curveto(9.165,0.435)(9.35,0.515)(9.42,0.58)
\curveto(9.49,0.645)(9.65,0.775)(9.74,0.84)
\curveto(9.83,0.905)(10.085,1.0)(10.25,1.03)
\curveto(10.415,1.06)(10.68,1.02)(10.78,0.95)
\curveto(10.88,0.88)(11.09,0.68)(11.2,0.55)
\curveto(11.31,0.42)(11.595,0.235)(11.77,0.18)
\curveto(11.945,0.125)(12.285,0.14)(12.45,0.21)
\curveto(12.615,0.28)(12.92,0.46)(13.06,0.57)
\curveto(13.2,0.68)(13.545,0.845)(13.75,0.9)
\curveto(13.955,0.955)(14.31,0.995)(14.76,0.95)
}
\psline[linewidth=0.04cm,linestyle=dashed,dash=0.16cm 0.16cm,arrowsize=0.05291667cm 2.0,arrowlength=1.4,arrowinset=0.4]{<-}(6.92,2.81)(6.92,-2.81)
\psline[linewidth=0.04cm,linestyle=dashed,dash=0.16cm 0.16cm,arrowsize=0.05291667cm 2.0,arrowlength=1.4,arrowinset=0.4]{->}(0.0,0.77)(15.58,0.77)
\usefont{T1}{ptm}{m}{n}
\rput(15.384531,0.44){$\x$}
\usefont{T1}{ptm}{m}{n}
\rput(6.4045315,2.6){$y$}
\usefont{T1}{ptm}{m}{n}
\rput(7.3745313,0.44){$O$}
\usefont{T1}{ptm}{m}{n}
\rput(4.2145314,1.48){$y=\eta(\x,t)$}
\usefont{T1}{ptm}{m}{n}
\rput(3.2345312,-1.34){$\Delta\phi = 0$}
\usefont{T1}{ptm}{m}{n}
\rput(11.104531,-2.3){$|\/\operatorname{grad}\/\phi\/|\to 0\quad\mathrm{as}\quad y\to -\infty$}
\end{pspicture}
}
\end{center}
\caption{\small\em Definition sketch of the fluid domain.}
\label{fig:sketch}
\end{figure}
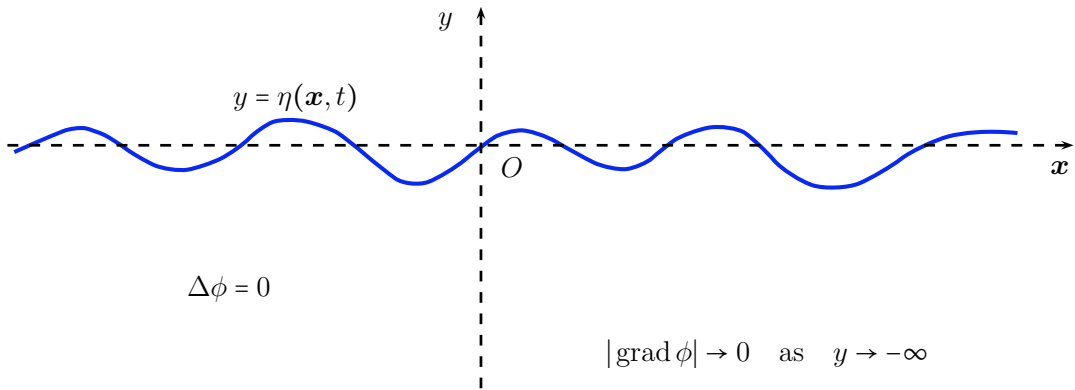

Consider an ideal incompressible fluid of constant density $\rho$. The vertical projection of the fluid domain $\Omega$ is a subset of $\R^2$. The horizontal independent variables are denoted by $\x = (x_1,x_2)$ and the upward vertical one by $y$. The origin of the Cartesian coordinate system is chosen such that the surface $y=0$ corresponds to the still water level. The fluid is bounded above by an impermeable free surface at $y=\eta(\vx,t)$. We assume that the fluid is unbounded below. This assumption constitutes the so-called deep water limiting case which is valid if the typical wavelength is much smaller than the average water depth. The sketch of the physical domain is shown in Figure~\ref{fig:sketch}.

\subsection{Fundamental equations}

Assuming that the flow is incompressible and irrotational, the governing equations of the classical water wave problem over an infinite depth are the following \cite{Lamb1932, Mei1994, Stoker1958, 
Whitham1999}:
\begin{align}
  \grad^2\phi\ +\ \dy^{\,2}\,\phi\ &=\ 0 \qquad
  -\infty\,\leqslant y\leqslant\eta(\x,t), \label{eq:laplace} \\
  \dt\eta\ +\ (\grad\phi)\scal(\grad\eta)\ -\ \dy\,\phi\ &=\ 0 
  \qquad y = \eta(\x, t), \label{eq:kinematic} \\
  \dt\phi\ +\ \half\,(\grad\phi)^2\ +\ \half\,(\dy\phi)^2\ +\ g\,\eta\ &=\ 0 
  \qquad y = \eta(\x,t), \label{eq:bernoulli} \\
  |\/\operatorname{grad}\/\phi\/|\ &\to\ 0 
  \qquad y \to -\infty, \label{eq:bottomkin}
\end{align}
with $\phi$ being the velocity potential (\ie, $\vu=\nab\phi$, $v=\dy\phi$), $g$ the acceleration due to gravity and where $\nab = (\partial_{x_1},\partial_{x_2})$ denotes the gradient operator in horizontal plane.

The incompressibility condition leads to the Laplace equation for $\phi$. The main difficulty of the water wave problem lies on the nonlinear free boundary conditions and that the free surface shape is unknown. Equation \eqref{eq:kinematic} expresses the free-surface kinematic condition, while the dynamic condition \eqref{eq:bernoulli} expresses the free surface isobarity. Finally, the last condition \eqref{eq:bottomkin} means that the velocity field decays to zero as $y \to -\infty$.

The water wave problem possesses several variational structures \cite{Luke1967,Petrov1964, Whitham1965, Zakharov1968}. In the present study, we will focus mainly on the Lagrangian variational formalism, but not exclusively. Surface gravity wave equations \eqref{eq:laplace}-\eqref{eq:bernoulli} can be derived as Euler--Lagrange equations of a functional proposed by Luke \cite{Luke1967}
\begin{equation}\label{eq:L0}
  \mathcal{L}\,=\int_{t_1}^{t_2}\!\int_{\Omega}\mathscr{L}\,\rho\ \ud^2\/\x\,\ud\/t, \quad
  \mathscr{L}\,=\,-\int_{-\infty}^\eta\left[\,g\,y\, +\, \dt\,\phi\, +\, \half\,(\grad\,\phi)^2\, +\, \half\,(\dy\,\phi)^2\,\right]\,\ud\/y.
\end{equation}
In a recent study, \textsc{Clamond} \& \textsc{Dutykh} \cite{Clamond2009} proposed to use Luke's Lagrangian \eqref{eq:L0} in the following relaxed form
\begin{align}\label{eq:L3}
\mathscr{L}\ =&\ (\eta_t+\vmus\scal\nabla\eta-\nus)\,\phis\ -\ \half\,g\,\eta^2\  \nonumber \\
  &+\ \int_{-\infty}^{\,\eta}\left[\,\vmu\scal\u-{\half}\/\u^2\,+\,\nu\/v-\half\/v^2\, +\,(\nabla\scal\vmu+\nu_y)\,\phi\,\right]\ud\/y,
\end{align}
where \{$\vu,v,\vmu,\nu$\} are the horizontal velocities, the vertical velocity and the associated Lagrange multipliers, respectively. The additional variables \{$\vmu,\nu$\} (Lagrange multipliers) are called pseudo-velocities. The over `tildes' denote a quantity computed at the free surface $y = \eta(\vx,t)$.

While the original Lagrangian \eqref{eq:L0} incorporates only two variables ($\eta$ and $\phi$), the relaxed Lagrangian density \eqref{eq:L3} involves six variables \{$\eta,\phi,\vu,v,\vmu,\nu$\}. These additional degrees of freedom provide us with more flexibility in constructing various approximations. For more details, explanations and examples we refer to \cite{Clamond2009}.

\subsection{Model equations}

Now, we illustrate the practical use of the variational principle \eqref{eq:L3} on an example borrowed from \cite{Clamond2009}. For progressive waves in deep water, the Stokes expansion shows that the velocity field varies nearly exponentially along the vertical. Even for very large unsteady waves (including breaking waves), accurate numerical simulations and experiments have shown that the vertical variation of the velocity field is indeed very close to an exponential \cite{Grue2003, Jensen2007}. Thus, this property is exploited here to derive a simple approximation for waves in deep water.

Let $\kappa > 0$ be a characteristic wavenumber corresponding, for example, to the carrier wave of a modulated wave group or to the peak frequency of a JONSWAP spectrum. Following the discussion above, it is natural to seek approximations in the form
\begin{equation}\label{ansinf}
\{\,\phi\,;\,\u\,;\,v\,;\,\vmu\,;\,\nu\,\}\ \approx\ \{\,\phis\,;\,\vus\,;\,\vs\,;\,\vmus\,;\,\nus\,\}\
\ue^{\kappa\/(y-\eta)},
\end{equation}
where $\phis$, $\vus$, $\vs$, $\vmus$ and $\nus$ are functions of $\vx$ and $t$ that will be determined using the variational principle. The ansatz \eqref{ansinf} iscertainly the simplest possible that is consistent with experimental evidences. This ansatz has already been used by \textsc{Kraenkel} \etal \cite{Kraenkel2005} for building their approximation.

For the sake of simplicity, we introduce the constraints $\vmus=\vus$ and $\nus=\vs$. Thus, the ansatz \eqref{ansinf} substituted into the Lagrangian density \eqref{eq:L3} yields
\begin{equation}\label{defL21}
  2\/\kappa\,\mathscr{L}\ =\ 2\/\kappa\, \phis\,\dt\eta\, - \,g\/\kappa\,\eta^2\, + \,\half\/\vus^2\, + \,\half\/\vs^2\, - \,\vus\scal(\grad\phis-\kappa\/\phis\grad\eta)\, - \,\kappa\,\vs\,\phis.
\end{equation}

The Euler--Lagrange equations for this functional yield
\begin{eqnarray*}
\delta\,\vus\/:&& 0\ =\ \vus\,-\,\nab\phis\,+\,\kappa\,\phis\,\nab\eta, \\
\delta\,\vs\/:&&  0\ =\ \vs\,-\,\kappa\,\phis, \\
\delta\,\phis\/:&& 0\ =\ 2\/\kappa\,\dt\eta\,+\,\nab\scal\vus\,-\,\kappa\,\vs\,+\,\kappa\,\vus\scal\nab\eta, \\
\delta\,\eta\/:&& 0\ =\
2\/g\/\kappa\,\eta\,+\,2\/\kappa\,\dt\phis\,+\,\kappa\,\nab\scal(\,\phis\,\vus\,).
\end{eqnarray*}
The two first relations imply that this approximation is exactly irrotational and their use in the last two equations gives
\begin{eqnarray}
& \dt\eta\,+\,\half\/\kappa^{-1}\nab^2\phis\,-\,\half\/\kappa\/\phis\ =\ \half\,\phis\left[\,\nab^2\eta\,+\,\kappa\,(\nab\eta)^2\,\right],& \label{eq:gkg1}\\
& \dt\phis\,+\,g\,\eta\ =\ -\half\,\nab\scal\left[\,\phis\,\nab\phis\,-\,\kappa\,\phis^2\, \nab\eta\,\right].& \label{eq:gkg2}
\end{eqnarray}

Since equations \eqref{eq:gkg1}--\eqref{eq:gkg2} derive from an irrotational motion, they can also be obtained from Luke's Lagrangian \eqref{eq:L0} under ansatz \eqref{ansinf}. However, we prefer to keep the heavy machinery of the relaxed variational principle since it allows to derive new models which cannot be obtained from Luke's variational principle. These examples will be investigated in future works. Equations \eqref{eq:gkg1}--\eqref{eq:gkg2} physically are a deep water counterpart of Saint-Venant equations for shallow water waves, in the sense given in \cite{Clamond2009}.

\begin{remark}
To the linear approximation, after elimination of $\phis$, equations \eqref{eq:gkg1}--\eqref{eq:gkg2} yield
\begin{equation}\label{gkglin}
  \dt^{\,2}\/\eta\ -\ \half\,(g/\kappa)\,\nab^2\eta\ +\ \half\,g\,\kappa\,\eta\ =\ 0,
\end{equation}
that is a Klein--Gordon equation. For this reason, equations \eqref{eq:gkg1}--\eqref{eq:gkg2} are referred here as ``\acf{gkg}''. The Klein--Gordon equation is prominent in mathematical physics and appears, e.g., as a relativistic generalisation of the Schr\"{o}dinger equation. The Klein--Gordon equation \eqref{gkglin} admits a special ($2\pi/k$)-periodic traveling wave solution
\begin{equation*}
  \eta\ =\ a\,\cos k\/(x_1-c\/t), \qquad c^2\ =\ \half\,g\,(k^2+\kappa^2)\,(\kappa\,k^2)^{-1}.
\end{equation*}
Therefore, if $k=\kappa$ the exact dispersion relation of linear waves (\ie, $c^2=g/k$) is recovered, as it should be. This means, in particular, that the \acs{gkg} model is valid for spectra narrow-banded around the wavenumber $\kappa$. 
\end{remark}

\section{Symplectic structures}\label{sec:struct}

In this section we unveil two other variational structures of the \acs{gkg} equations.

\subsection{Canonical Hamiltonian}

It is straightforward to verify that the \acs{gkg} equations possess a canonical Hamiltonian structure
\begin{equation*}
  \begin{pmatrix}
    \dt\eta \\
    \dt\phis
  \end{pmatrix}\ =\ \J\scal\begin{pmatrix}
                     \displaystyle{\delta\,\H\,/\,\delta\/ \phis} \\
                     \displaystyle{\delta\,\H\,/\,\delta\/ \eta}
                   \end{pmatrix},
  \qquad
  \J = \begin{pmatrix}
    0 & -1 \\
    1 & 0
  \end{pmatrix},
\end{equation*}
with the Hamiltonian functional $\H$ is defined as
\begin{equation}\label{Hgkg}
\H = \int_{\Omega}\left\{\,\half\,g\,\eta^2\,+\,\quat\,\kappa^{-1}\left[\,\nab\phis\,-\,
\kappa\,\phis\,\nab\eta\,\right]^2\,+\,\quat\,\kappa\,{\phis}^2\,\right\}\,\ud^2\/\x.
\end{equation}
This `simple' Hamiltonian $\H$ is quartic in nonlinearities and involves only first-order derivatives. It has to be compared with Zakharov's quartic Hamiltonian which involves second-order derivatives and pseudo-differential operators. However, Zakharov's quartic Hamiltonian is valid for broad spectra, while the \acs{gkg} are limited to very narrow-banded spectra. Note that the Hamiltonian \eqref{Hgkg} {cannot be derived from the exact one}, since the latter assumes that irrotationality and incompressibility are both satisfied identically in the bulk, while the incompressibility is not fulfilled by equations \eqref{eq:gkg1}--\eqref{eq:gkg2}.

\subsection{Multi-symplectic structure}

In addition to the Lagrangian and Hamiltonian formulations, the \acs{gkg} equations \eqref{eq:gkg1}--\eqref{eq:gkg2} can be recast into a first-order PDE system:
\begin{align}
2\,\kappa\,\dt\/\eta\ +\ \nab\scal\vus\ &=\ \kappa^2\,\phis\ -\ \kappa\,\vus\scal\valpha, \\
-\/2\,\kappa\,\dt\/\phis\ -\ \nab\scal\vgamma\ &=\ 2\,\kappa\,g\,\eta, \\
-\/\nab\phis\ &=\ -\/\vus\ -\ \kappa\,\phis\,\valpha, \\
\nab\eta\ &=\ \valpha, \\
\boldsymbol{0}\ &=\  \vgamma\ -\ \kappa\,\phis\,\vus,
\end{align}
where $\valpha = (\alpha_1, \alpha_2)$ and $\vgamma = (\gamma_1, \gamma_2)$ are auxiliary variables. These relations yield the multi-symplectic canonical structure
\begin{equation}\label{eq:ms}
\M\cdot\vec{z}_t\ +\ \K_1\cdot\vec{z}_{x_1}\ +\ \K_2\cdot\vec{z}_{x_2}\ =\ 
\operatorname{grad}_{\vec{z}}\,\S(\vec{z}),
\end{equation}
where $\vec{z}=(\phis,\eta,\us_1,\us_2,\gamma_1,\gamma_2,\alpha_1,\alpha_2)^\text{T} \in \R^{8}$, 
$\S$ is the generalised Hamiltonian function 
\[
\S(\vec{z})\ =\ \valpha\scal\vgamma \ +\  \kappa\,g\,\eta^2\ +\  
\half\left(\kappa\/\phis\right)^2 \ -\ \kappa\,\phis\,\vus\scal\valpha\ -\  
\half\,\vus\scal\vus,
\]
and where the eight-by-eight skew-symmetric matrices $\M$, $\K_1$ and $\K_2$ are defined as 
\begin{align}
\M\ &=\ 2\,\kappa\left(\vec{e}_1\otimes\vec{e}_2\,-\,\vec{e}_2\otimes\vec{e}_1\right),\\
\K_1\ &=\ \vec{e}_1\otimes\vec{e}_3\ -\ \vec{e}_3\otimes\vec{e}_1\ +\ 
\vec{e}_5\otimes\vec{e}_2-\ \vec{e}_2\otimes\vec{e}_5,\\
\K_2\ &=\ \vec{e}_1\otimes\vec{e}_4\ -\ \vec{e}_4\otimes\vec{e}_1\ +\ 
\vec{e}_6\otimes\vec{e}_2-\ \vec{e}_2\otimes\vec{e}_6,
\end{align}
$\vec{e}_j$ being $j$-th unitary vector of the Cartesian coordinates for the $\R^{8}$ space ($\otimes$ the tensor product).

\subsection{Conservation laws}

The local multi-symplectic conservation law for \eqref{eq:ms} is
\begin{eqnarray*}
  \partial_t\,\omega\ +\ \nab\scal\vtau\ =\ 0,
\end{eqnarray*}
where the pre-symplectic forms are defined, for any solution of the first variation of \eqref{eq:ms}, as
\begin{align*}
\omega\ =\ \half\, \ud \vec{z} \wedge (\/\M\cdot\ud\vec{z}\/), \qquad
\tau_1\ =\ \half\, \ud \vec{z} \wedge (\/\K_1\cdot\ud\vec{z}\/), \qquad
\tau_2\ =\ \half\, \ud \vec{z} \wedge (\/\K_2\cdot\ud\vec{z}\/), 
\end{align*}
that is to say
\begin{align*}
\omega\, =\, 2\,\kappa\,\ud\/\eta \wedge \ud\/\phis, \quad
\tau_1\, =\, \ud \us_1 \wedge \ud \phis\, +\, \ud \gamma_1 \wedge \ud \eta, \quad
\tau_2\, =\, \ud \us_2 \wedge \ud \phis\, +\, \ud \eta \wedge \ud \gamma_2,
\end{align*}
where $\wedge$ is the usual exterior or wedge product \cite{Bridges2001,Olver1993}.

Along the multi-symplectic system solutions, local energy conservation law is verified
\begin{eqnarray*}
  \dt\,E(\/\vec{z}\/)\ +\ \nab\scal\vF(\/\vec{z}\/)\ =\ 0,
\end{eqnarray*}
with
\begin{align*}
  E\ &=\ \S\ -\ \half\,{\vec{z}}^{\,\text{T}}\cdot\K_1\cdot\vec{z}_{x_1}\ 
-\ \half\,{\vec{z}}^{\,\text{T}}\cdot\K_2\cdot\vec{z}_{x_2}, \qquad
  F_j\ =\ \half\,{\vec{z}}^{\,\text{T}}\cdot\K_j\cdot\vec{z}_t,
\end{align*}
which can be explicitly expressed in terms of the physical variables as
\begin{align*}
  2\,E\ &=\ 2\,\kappa\,g\,\eta^2\ -\ \vus^2\ +\ (\kappa\/\phis)^2\ -\ 
\kappa\,\phis\,\vus\scal\nab\eta\ +\ \kappa\,\eta\,\nab\scal(\phis\/\vus)\ -\ 
\phis\,\nab\scal\vus\ +\ \vus\scal\nab\phis,  \\
  2\,\vF\ &=\  \kappa\,\phis\,\vus\,\dt\eta\ -\ \kappa\,\eta\,\dt(\phis\/\vus)\ 
+\ \phis\,\dt\vus\ -\ \vus\,\dt\phis.    
\end{align*}
There exists also two local momentum conservation laws associated to each spatial direction 
\begin{align*}
\dt\,I_1(\vec{z})\ +\ \partial_{x_1}G_{11}(\vec{z})\ +\ \partial_{x_2}G_{12}(\vec{z})\ &=\ 0, \\
\dt\,I_2(\vec{z})\ +\ \partial_{x_1}G_{21}(\vec{z})\ +\ \partial_{x_2}G_{22}(\vec{z})\ &=\ 0,
\end{align*}
the corresponding quantities being
\begin{align*}
  2\,I_j\ &=\ {\vec{z}}^{\,\text{T}}\cdot\M\cdot\vec{z}_{x_j}\ =\ 
2 \kappa \left( \phis\,\partial_{x_j}\eta\ -\ \eta\,\partial_{x_j}\phis \right), \\
  2\,G_{12}\ &=\ \vec{z}^{\,\text{T}}\cdot\K_2\cdot\vec{z}_{x_1}\ =\ 
\kappa\,\phis\,\us_2\,\partial_{x_1}\eta\ -\ \kappa\,\eta\,\partial_{x_1}(\phis\/\us_2)\  
+\ \phis\,\,\partial_{x_1}\us_2\ -\ \us_2\,\,\partial_{x_1}\phis, \\
  2\,G_{21}\ &=\ \vec{z}^{\,\text{T}}\cdot\K_1\cdot\vec{z}_{x_2}\ =\ 
\kappa\,\phis\,\us_1\,\partial_{x_2}\eta\ -\ \kappa\,\eta\,\partial_{x_2}(\phis\/\us_1)\  
+\ \phis\,\,\partial_{x_2}\us_1\ -\ \us_1\,\,\partial_{x_2}\phis, \\
  2\,G_{11}\ &=\ 2\,\S\ -\ \vec{z}^{\,\text{T}}\cdot\M\cdot\vec{z}_t\ -\ 
\vec{z}^{\,\text{T}}\cdot\K_2\cdot\vec{z}_{x_2} \\ 
&=\ 2\,\kappa\,g\,\eta^2\ +\ (\/\kappa\/\phis\/)^2\ -\ \vus^2\ +\ 2 \kappa \left(\eta\,\dt\phis\ 
-\ \phis\,\dt\eta\ \right) \\ 
&\quad -\,\left(\kappa\,\phis\,\us_2\,\partial_{x_2}\eta\ -\ \kappa\,\eta\,\partial_{x_2}
(\phis\,\us_2)\ +\ \phis\,\,\partial_{x_2}\us_2\ -\ \us_2\,\,\partial_{x_2}\phis \right), \\
2\,G_{22}\ &=\ 2\,\S\ -\ \vec{z}^{\,\text{T}}\cdot\M\cdot\vec{z}_t\ -\ 
\vec{z}^{\,\text{T}}\cdot\K_1\cdot\vec{z}_{x_1}  \\ 
&=\ 2\,\kappa\,g\,\eta^2\ +\ (\/\kappa\/\phis\/)^2\ -\ \vus^2\ +\ 2 \kappa \left(\eta\,\dt\phis\ 
-\ \phis\,\dt\eta\ \right) \\ 
&\quad -\,\left(\kappa\,\phis\,\us_1\,\partial_{x_1}\eta\ -\ \kappa\,\eta\,\partial_{x_1}
(\phis\,\us_1)\  +\ \phis\,\,\partial_{x_1}\us_1\ -\ \us_1\,\,\partial_{x_1}\phis\right). 
\end{align*}

The multi-symplectic form highlighted above can be used to construct various numerical multi-symplectic schemes which preserve exactly the multi-symplectic form at the discrete level \cite{Bridges2001, Moore2003, Schober2008, Zhao2000, Bridges2006}. These schemes are suitable for the long time integration using rather coarse discretizations \cite{Dutykh2013a}. Since in the present manuscript we focus on mid-range, but highly accurate simulations we opt for the pseudo-spectral discretizations. However, the multi-symplectic framework seems to be very promising for long time dynamics investigations employing only a moderate number of the degrees of freedom. The conserved quantities derived above can be used to assess the accuracy of the numerical solution, even if their physical meaning is not clear at the current stage.

\section{Travelling waves}\label{sec:tw}

For the sake of simplicity we will consider hereinafter the special case of two-dimensional wave motions, \ie the dependent variables are independent of, say, the variable $x_2$; for brevity, we denote $x=x_1$ and $u = u_1$. The equations of motions become
\begin{align}
  \us\ &=\ \phis_x\ -\ \kappa\,\phis\,\eta_x, \label{equ2D} \\
  \vs\ &=\ \kappa\,\phis, \label{eqv2D} \\
0\ &=\ 2\,\kappa\,\eta_t\ +\ \us_x\ -\ \kappa\,\vs\ +\ \kappa\,\us\,\eta_x, \label{eqet2D}\\
  0\ &=\ 2\,g\,\kappa\,\eta\ +\ 2\,\kappa\,\phis_t\ +\ [\,\us\,\vs\,]_x \label{eqft2D},
\end{align}
which can be reduced into a two equations system
\begin{align}
  \eta_t\ +\ \half\,\kappa^{-1}\,\phis_{xx}\ -\ \half\,\kappa\,\phis\ &=\ \half\,\phis
\left[\,\eta_{xx}\,+\,\kappa\,\eta_x^{\,2}\,\right], \label{eq2D:gkg1}\\
  \phis_t\ +\ g\,\eta\ &=\ -\half\left[\,\phis\,\phis_x\,-\,\kappa\,\phis^2\, 
\eta_x\,\right]_x. \label{eq2D:gkg2}
\end{align}

The equations can be combined to derive useful secondary relations. For instance, we derive the conservative equations
\begin{align}
  \tilde{u}_t\ +\, \left[\, {\textstyle{3\over4}}\, \tilde{u}^2\, + \, 
  {\textstyle{1\over4}}\, \tilde{v}^2\, + \,g\,\eta \left(1 - 
  \half\kappa\/\eta\right)\,\right]_x\ &=\ 0, \label{eqmonflux} \\
\left[\,\half\,g\,\kappa\,\eta^2\,+\,{\textstyle{1\over4}}\,(\tilde{u}^2+\tilde{v}^2)\,\right]_t\ 
+\, \left[\, \half\,\tilde{u}\, (\tilde{v}\,\eta_t\,-\,\tilde{\phi}_t)\, 
\right]_x\ &=\ 0, \label{eqeneflux}
\end{align}
which physically describe (after division by $\kappa$) the conservations of the momentum and energy fluxes, respectively.

For traveling waves of permanent form, the dependent variables are functions of the single independent variable $\xi = x - ct$. The equations \eqref{eqmonflux} and \eqref{eqeneflux} can then be integrated as
\begin{align}
  {\textstyle{3\over4}}\,\tilde{u}^2\ +\ {\textstyle{1\over4}}\,
  \tilde{v}^2\ +\ g\,\eta\left(1-\half\kappa\/\eta\right)\ -\ c\,\tilde{u}\ &=\ K_p, \label{eqmonfluxpro} \\
  \half\,g\,\kappa\,\eta^2\ -\ {\textstyle{1\over4}}\,\tilde{u}^2\ +\ {\textstyle{1\over4}}\,\tilde{v}^2\ &=\ K_e, \label{eqenefluxpro}
\end{align}
where $K_p$ and $K_e$ are integration constants. Adding these two relations, one obtains
\begin{equation}
  \half\,\tilde{u}^2\ +\ \half\,\tilde{v}^2\ +\ g\,\eta\ -\ c\,\tilde{u}\ =\ K_p\ +\ K_e,
\end{equation}
which is the Bernoulli equation, $K_p + K_e$ being a Bernoulli constant. Subtracting the two relations, one gets at once
\begin{equation}
  \tilde{u}^2\ +\ g\,\eta\left(1-\kappa\/\eta\right)\ -\ c\,\tilde{u}\ =\ K_p\ -\ K_e,
\end{equation}
that can be used to express $\tilde{u}$ in terms of $\eta$ (or vice-versa), \ie,
\begin{align}
  \tilde{u}\ &=\ \half\,c\ \pm\ \sqrt{\,K_p\,-\,K_e\,+\,{\textstyle{1\over4}}\,c^2\, - \,g\,\eta\left(1-\kappa\/\eta\right)\,}, \label{soluseta} \\ \tilde{u}_\xi\ &=\ g\,\eta_\xi\,(\/1\/-\/2\/\kappa\/\eta\/)\,/\,(\/c\/-\/2\/\us\/). 
\end{align}

With these relations, the Lagrangian density \eqref{defL21} becomes
\begin{align}\label{defL2Dpro}
  2\/\kappa\,\mathscr{L}\ &=\ -\/2\,c\,\vs\,\eta_\xi\ -\ g\,\kappa\,\eta^2\ -\ \half\,\us^2\ -\ \half\,\vs^2 \nonumber \\ 
  &=\ 2\,c\,\eta_\xi^{\,2}\left[\,2\,c\,-\,\us\,-\,(g/\kappa)\,(\/1\/-\/2\/\kappa\/\eta\/)\,/\,(\/c\/-\/2\/\us\/)\,\right]\, -\ \us^2\ -\ 2\,K_e,
\end{align}
where $\us$ should be expressed via \eqref{soluseta}. An equation for $\eta$ is then obtained from the Beltrami identity
\[
  \mathscr{L}\ -\ \eta_\xi\,\frac{\partial\,\mathscr{L}}{\partial\/\eta_\xi}\ =\ \text{constant}\ \equiv\ (\/K_b\/-\/2\/K_e\/)\,/\,2\/\kappa,
\]
yielding
\begin{equation}\label{odeprog}
  \left(\frac{\ud\,\eta}{\ud\/\xi}\right)^{\!2}\ =\ \frac{\kappa\,(\/K_b\/+\/\us^2\/) \,(\/c\/-\/2\/\us\/)}{2\,c\,[\,g\,(\/1\/-\/2\/\kappa\/\eta\/)\,+\,\kappa\,(\/2\/c\/-\/\us\/)\,(\/c\/-\/2\/\us\/)\,]},
\end{equation}
where $\us$ is given by \eqref{soluseta}. Unfortunately, we were not able to solve equation \eqref{odeprog} analytically. However, this solution might be useful for theoretical investigations of travelling waves. In order to construct these solutions numerically to high accuracy ($\sim 10^{-10}$), we employ a Newton Jacobian-free method combined with the Levenberg--Marquardt algorithm \cite{More1978}. The computed profiles will be shown below in Section~\ref{sec:numres}.

\subsection{Stokes wave}\label{sec:compa}

Despite the fact that we were not able to find exact analytical solutions to the \acs{gkg} equations, a Stokes-type expansion can help us to evaluate the accuracy of the approximate model. To the seventh-order, a asymptotic expansion of the ($2\pi/\kappa$)-periodic progressive wave of \acs{gkg} equations is 
\cite{Clamond2009}:
\begin{eqnarray*}
\kappa\,\eta &=& \alpha\cos\xi
 \, + \,{\textstyle{1\over2}}\alpha^2\!\left(1\! + \!{\textstyle{{\bf25}\over12}}\alpha^2\!
 + \!{\textstyle{{\bf1675}\over{\bf192}}}\alpha^4\right)\cos2\xi \nonumber \\
&& + \,{\textstyle{3\over8}}\alpha^3\!\left(1\! + \!
{\textstyle{{\bf99}\over16}}\alpha^2\! + \!{\textstyle{{\bf11807}\over320}}\alpha^4\right)
\cos3\xi\, + \,{\textstyle{1\over3}}\alpha^4\!\left(1\! + \!
{\textstyle{{\bf64}\over{\bf5}}}\alpha^2\right)\cos4\xi \\
&& + \,{\textstyle{125\over384}}\alpha^5\!\left(1\! + \!
{\textstyle{{\bf6797}\over{\bf300}}}\alpha^2\right)\cos5\xi
\, + \,{\textstyle{27\over80}}\alpha^6\cos6\xi
\, + \,{\textstyle{16807\over46080}}\alpha^7\cos7\xi\, +\, \O(\alpha^8), \nonumber \\
{g^{-{1\over2}}}\/\kappa^{3\over2}\,\phis &=& \alpha\left(1\!-\!{\textstyle{1\over4}}\alpha^2\! 
-\!{\textstyle{{\bf59}\over96}}\alpha^4\! - \!{\textstyle{{\bf4741}\over1536}}\alpha^6\right)
\sin\xi\, + \, {\textstyle{1\over2}}\alpha^2\!\left(1\!+\!{\textstyle{{\bf11}\over12}}\alpha^2\!
 + \!{\textstyle{{\bf547}\over{\bf192}}}\alpha^4\right)\sin2\xi \nonumber \\
&& + \,{\textstyle{3\over8}}\alpha^3\!\left(1\! + \!
{\textstyle{{\bf163}\over{\bf48}}}\alpha^2\! + \!{\textstyle{{\bf221}\over{\bf15}}}\alpha^4\right)
\sin3\xi\, + \,{\textstyle{1\over3}}\alpha^4\!\left(1\! + 
\!{\textstyle{{\bf149}\over{\bf20}}}\alpha^2\right)\sin4\xi \\
&&+\,{\textstyle{125\over384}}\alpha^5\!\left(1\! + \! {\textstyle{{\bf5057}\over{\bf375}}}
\alpha^2\right)\sin5\xi\, + \,{\textstyle{27\over80}}\alpha^6\sin6\xi\, + \,
{\textstyle{16807\over46080}}\alpha^7\sin7\xi\, + \,\O(\alpha^8), \nonumber \\
{g^{-{1\over2}}}\/\kappa^{1\over2}\,c &=& 1 + {\textstyle{1\over2}}\alpha^2
 + {\textstyle{1\over2}}\alpha^4 + {\textstyle{{\bf899}\over384}}\alpha^6\,+\,\O(\alpha^8).
\end{eqnarray*}
For comparison, the corresponding seventh-order Stokes expansion of the full Euler equations in deep 
water is given by the following formulas
\begin{eqnarray}\label{eq:stokes}
\kappa\,\eta &=& \alpha\cos\xi\, + \,{\textstyle{1\over2}}\alpha^2\!\left(1\! + 
\!{\textstyle{17\over12}}\alpha^2\! + \!{\textstyle{233\over64}}\alpha^4\right)\cos2\xi 
\qquad \nonumber \\
&& + \,{\textstyle{3\over8}}\alpha^3\!\left(1\! + \!
{\textstyle{51\over16}}\alpha^2\!+\!{\textstyle{3463\over320}}\alpha^4\right)\cos3\xi
\, + \,{\textstyle{1\over3}}\alpha^4\!\left(1\! + \!
{\textstyle{307\over60}}\alpha^2\right)\cos4\xi \\
&& + \,{\textstyle{125\over384}}\alpha^5\!\left(1\! + \!
{\textstyle{10697\over1500}}\alpha^2\right)\cos5\xi\, + \,{\textstyle{27\over80}}\alpha^6
\cos6\xi\, + \,{\textstyle{16807\over46080}}\alpha^7\cos7\xi\, + \,\O(\alpha^8), \nonumber \\
{g^{-{1\over2}}}\/\kappa^{3\over2}\,\phis &=& \alpha\left(1\! - \!{\textstyle{1\over4}}\alpha^2\!
 - \!{\textstyle{43\over96}}\alpha^4\! - \!{\textstyle{2261\over1536}}\alpha^6\right)\sin\xi
 \, + \,{\textstyle{1\over2}}\alpha^2\!\left(1\! + \!{\textstyle{7\over12}}\alpha^2\!
 + \!{\textstyle{81\over64}}\alpha^4\right)\sin2\xi \nonumber \\
&& + \,{\textstyle{3\over8}}\alpha^3\!\left(1\! + \!
{\textstyle{281\over144}}\alpha^2\! + \!{\textstyle{5813\over1080}}\alpha^4\right)\sin3\xi
\, + \,{\textstyle{1\over3}}\alpha^4\!\left(1\! + \!
{\textstyle{431\over120}}\alpha^2\right)\sin4\xi  \\
&& + \,{\textstyle{125\over384}}\alpha^5\!\left(1\! + \!
{\textstyle{3369\over625}}\alpha^2\right)\sin5\xi
\, + \,{\textstyle{27\over80}}\alpha^6\sin6\xi
\, + \,{\textstyle{16807\over46080}}\alpha^7\sin7\xi\, + \,\O(\alpha^8), \nonumber \\
{g^{-{1\over2}}}\/\kappa^{1\over2}\,c &=& 1 + {\textstyle{1\over2}}\alpha^2
 + {\textstyle{1\over2}}\alpha^4 + {\textstyle{707\over384}}\alpha^6 + \O(\alpha^8), \label{eq:stokes3}
\end{eqnarray}
The expansions of $\eta$ and $\phis$ of the \acs{gkg} periodic solution match the exact Stokes wave up to the third-order (non-matching coefficients are displayed in bold). This is not surprising since the \acs{gkg} equations are cubic in nonlinearities. However, it is much more surprising is that the phase velocity $c$ is correct up to the fifth-order.

\section{Pseudo-spectral method}\label{sec:nums}

We briefly describe below a highly accurate Fourier-type pseudo-spectral method \cite{Boyd2000, Trefethen2000} that we use to simulate the dynamics of the \acs{gkg} equations. These methods have been proven to be extremely efficient (practically unbeatable) in the idealised periodic setting \cite{Boyd2000}. Below we show that the \acs{gkg} system can be integrated up to the Riemann wave breaking using the proposed pseudo-spectral scheme.

With $\uu=(\eta,\phis)^T$ denoting the vector of dynamic variables, the \acs{gkg} system 
\eqref{eq:gkg1}, \eqref{eq:gkg2} can be recast in the vector form
\begin{equation}\label{eq:oper}
  \uu_t\ +\ \Ll\scal\uu\ =\ N(\uu),
\end{equation}
where the operator $N$ denotes the right-hand side of equations \eqref{eq:gkg1}, \eqref{eq:gkg2} and the linear operator $\Ll$ is defined as
\begin{equation*}
  \Ll\ =\ \begin{bmatrix}
    \ 0\ & \frac{\grad^2 -\kappa^2}{2\kappa}\  \\
    \ g\ & 0\
  \end{bmatrix}, \qquad
  \hat\Ll\ =\ \begin{bmatrix}
   \ 0\ & -\frac{|\k|^2\/+\/\kappa^2}{2\kappa}\ \\
   \ g\ & 0\
  \end{bmatrix},
\end{equation*}
where $\hat\Ll$ is the operator $\Ll$ in Fourier space. The equation \eqref{eq:oper} is solved applying the Fourier transform in the spatial variable $\x$. The transformed variables is denoted by $\hat{\uu}(t,\k) = \F\{\uu(t,\x)\}$, $\k$ being the Fourier transform parameter. The nonlinear terms are computed in the physical space, while spatial derivatives are computed spectrally in the Fourier space. For example, the term $\phis\grad^2\eta$ is discretised as:
\begin{equation*}
  \F\left\{\,\phis\,\grad^2\eta\,\right\}\ =\ \F\left\{\,\F^{-1}\left(\hat{\phi}\right)
  \,\times\,\F^{-1}\{-|\k|^2\hat{\eta}\}\,\right\}.
\end{equation*}
The other nonlinear terms are treated in a similar way. We note that the usual three-half rule has 
to be applied for anti-aliasing \cite{Clamond2001, Fructus2005, Trefethen2000}.

In order to improve the stability of the time discretisation procedure, we integrate exactly the 
linear terms. This is achieved by making a change of variables \cite{Fructus2005, Milewski1999}:
\begin{equation}
\hat{\vv}(t)\ =\ \exp\!\left((t-t_0)\hat{\Ll}\right)\cdot\hat{\uu}(t), 
\qquad \hat{\vv}(t_0)\ =\ \hat{\uu}(t_0),
\end{equation}
yielding the equation
\begin{equation*}
\hat{\vv}_t\ =\ \exp\!\left((t-t_0)\hat{\Ll}\right)\cdot\F\left\{
N\!\left(\exp\!\left((t_0-t)\hat{\Ll}\right)\cdot\hat{\vv}\right)\right\}.
\end{equation*}
The exponential matrix of the operator $\hat{\Ll}$ can be explicitly computed to give
\begin{equation*}
  \exp\!\left((t-t_0)\hat{\Ll}\right)\, =\ \begin{bmatrix}
  \cos\bigl(\omega(t-t_0)\bigr) & -(\omega/g)\sin\bigl(\omega(t-t_0)\bigr) \\
  (g/\omega)\sin\bigl(\omega(t-t_0)\bigr) & \cos\bigl(\omega(t-t_0)\bigr)
  \end{bmatrix}, \qquad
  \omega^2\ =\ \frac{g\,\kappa}{2}\ +\ \frac{g\,|\k|^2}{2\,\kappa}.
\end{equation*}
Finally, the resulting system of ODEs is discretised in space by the Verner's embedded adaptive 9(8) Runge--Kutta scheme \cite{Verner1978}. The step size is chosen adaptively using the so-called H211b digital filter \cite{Soderlind2003, Soderlind2006} to meet the prescribed error tolerance, set as of the order of machine precision.

\section{Numerical results and discussion}\label{sec:numres}

\subsection{Periodic steady solutions}

We begin the numerical study of \acs{gkg} equations by computing numerically steady periodic Stokes-like solutions. We employ the Newton Levenberg--Marquardt method which tends to the steepest descent far from the solution (to ensure the convergence) and becomes the classical Newton method in the vicinity of the root \cite{More1978}. Then, we compare the computed profile to the seventh-order Stokes expansion to the full Euler equations \eqref{eq:stokes}--\eqref{eq:stokes3}. In order to fix the ideas, we choose the wavelength to be $\lambda \equiv 2\ell = 2\pi$, \ie the computational domain is $[-\ell, \ell]$. Consequently, the parameter $\kappa = {2\pi}/{\lambda} = 1$. For simplicity, we take also $g = 1$ $\mathsf{m}/\mathsf{s}^2$. In steady computations, we use only $N = 128$ Fourier modes. It is sufficient to compute to high accuracy ($\sim \O(10^{-9})$) the numerical solution at the collocation points.

\begin{figure}
  \centering
  \subfigure[$\eps = 0.1821$]{%
  \includegraphics[width=0.49\textwidth]{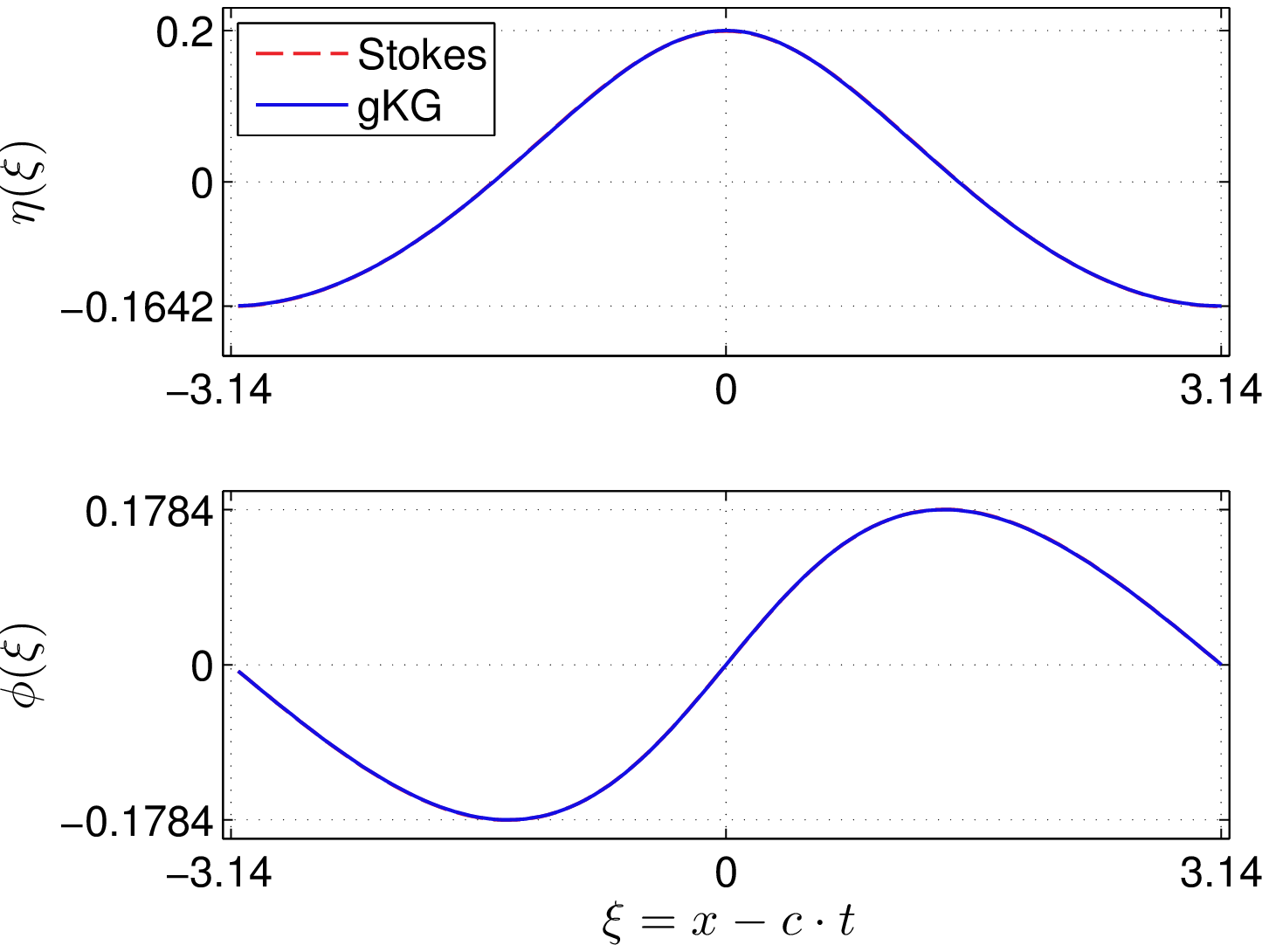}}
  \subfigure[$\eps = 0.2220$]{%
  \includegraphics[width=0.49\textwidth]{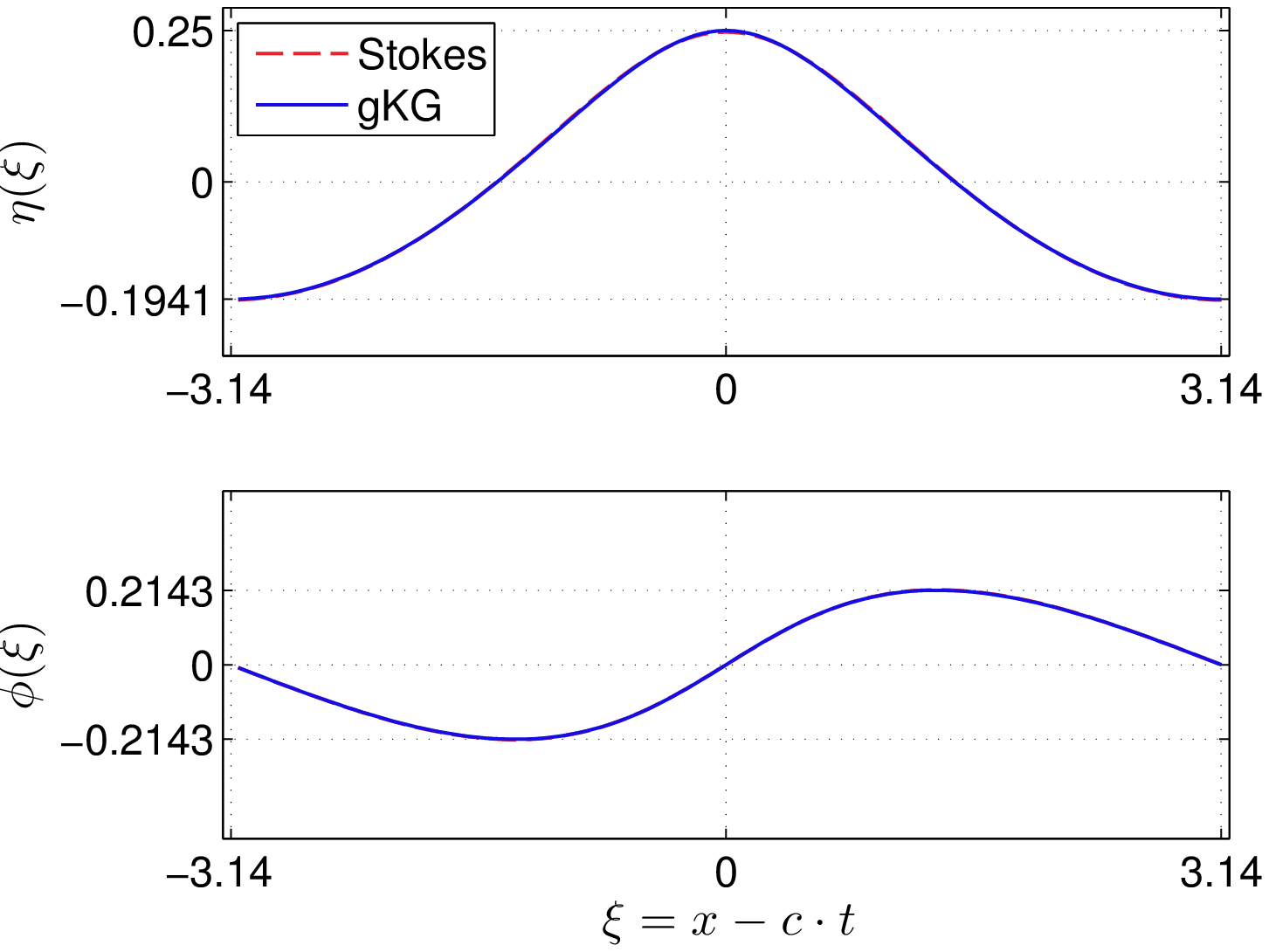}}
  \subfigure[$\eps = 0.2520$]{%
  \includegraphics[width=0.49\textwidth]{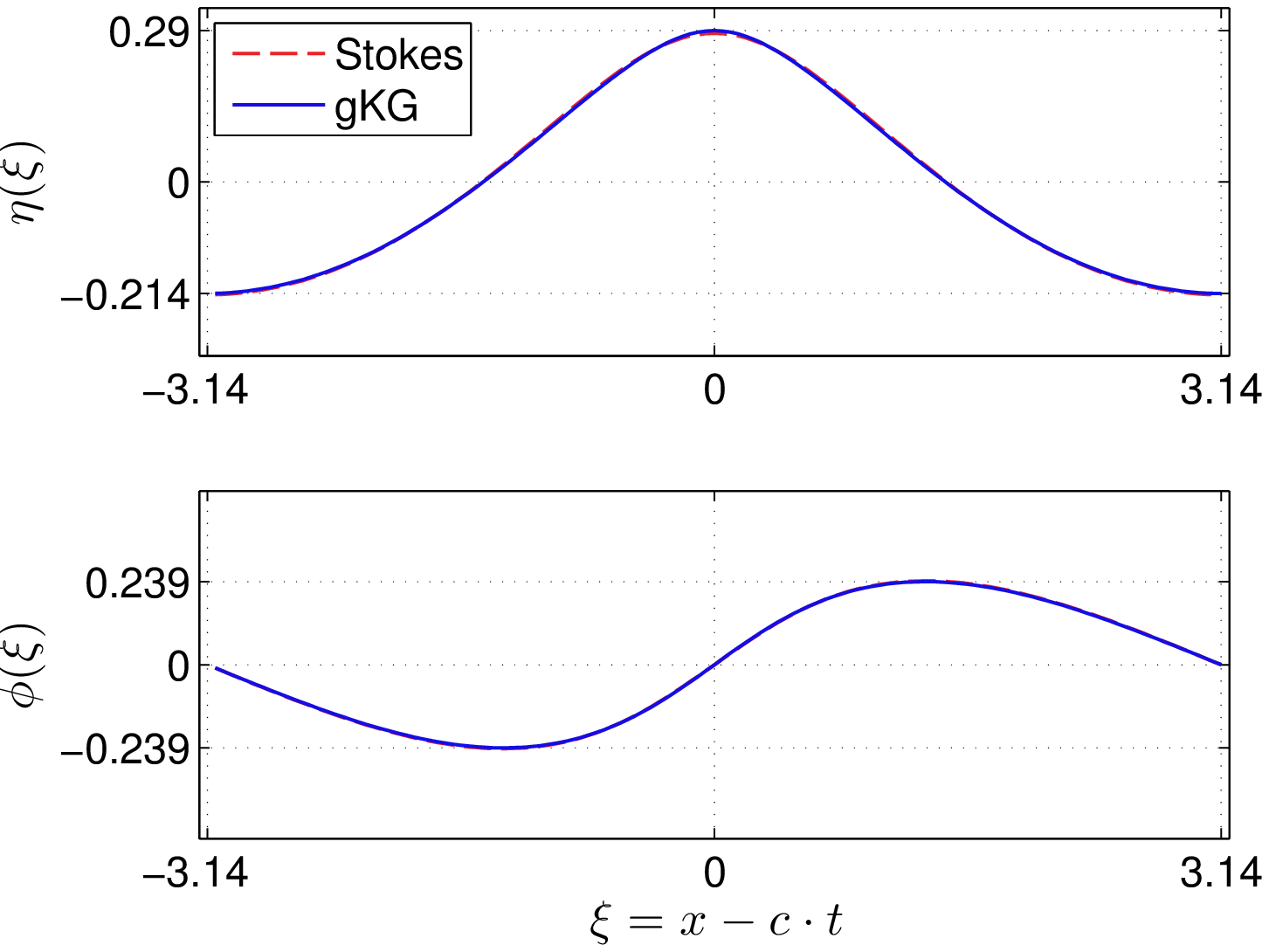}}
  \subfigure[$\eps = 0.2917$]{%
  \includegraphics[width=0.49\textwidth]{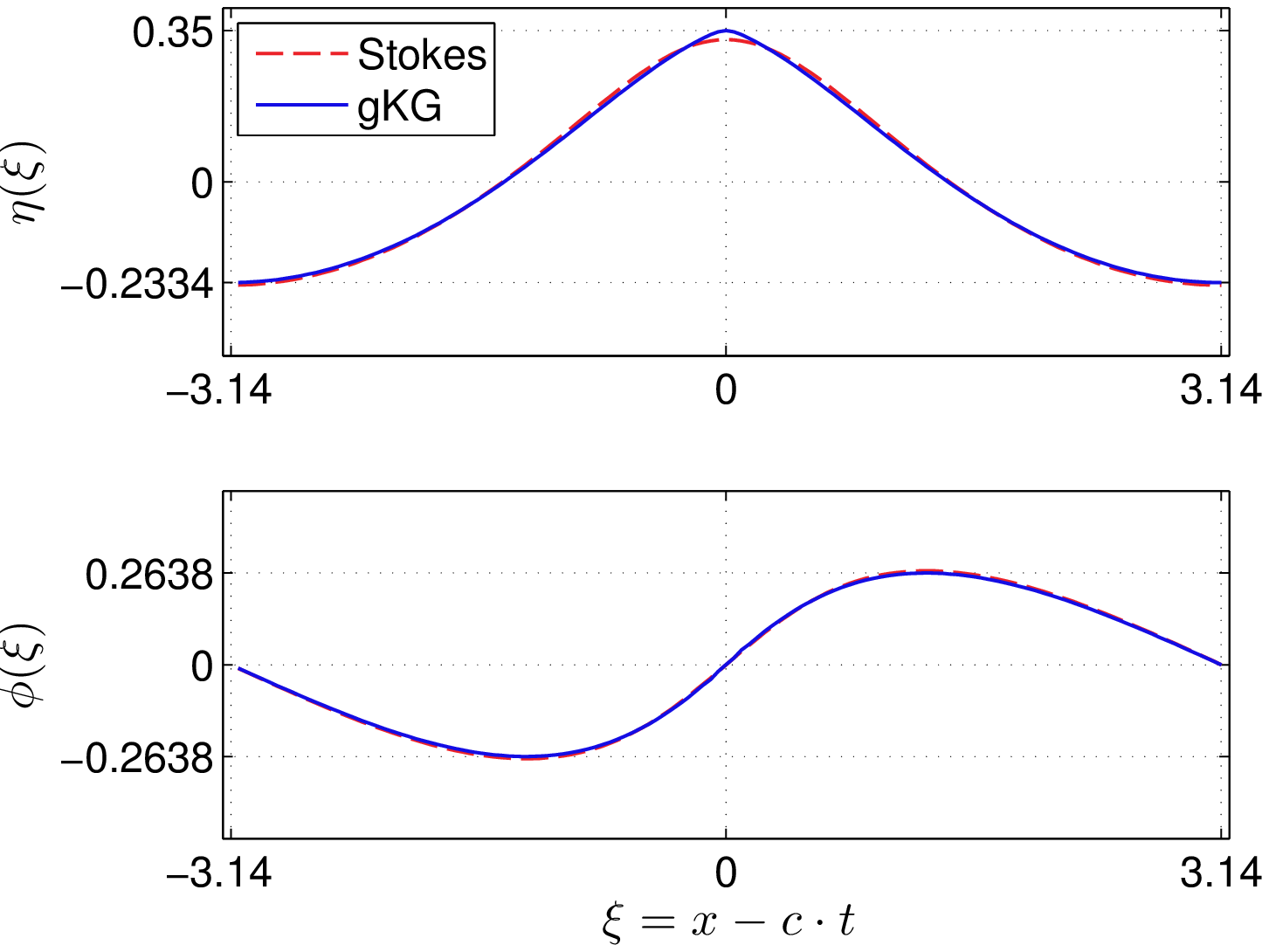}}
  \caption{\small\em Comparison of the travelling wave solutions to the \acs{gkg} equations with 
  the seventh-order Stokes solution for various values of the wave steepness parameter. The 
  wavelength is fixed to $2\pi$.}
  \label{fig:steady}
\end{figure}

\begin{figure}
  \centering
  \includegraphics[width=0.70\textwidth]{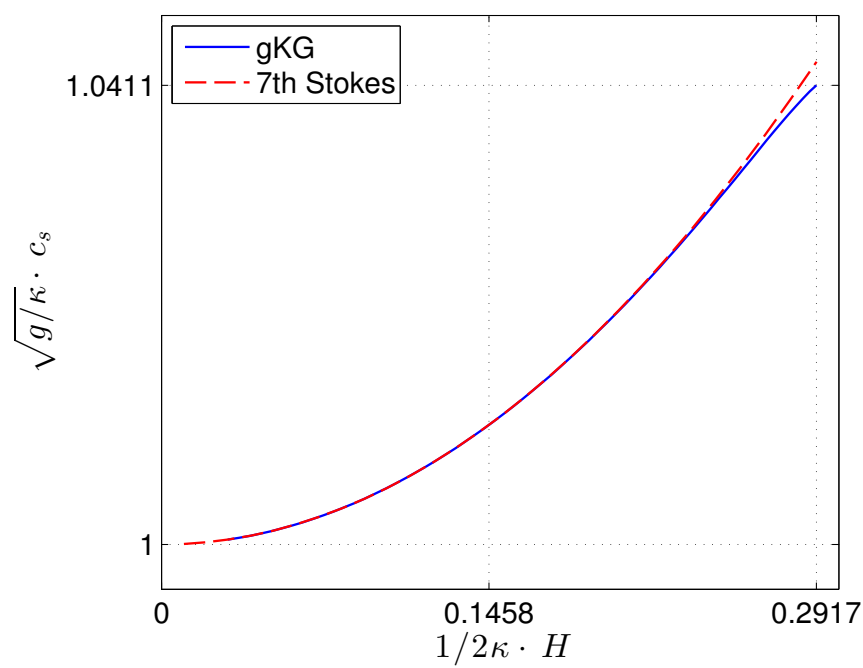}
  \caption{\small\em Speed--steepness relation for periodic steady waves: blue solid line --- 
  the \acs{gkg} equations, red dashed line --- seventh-order Stokes expansion. The wavelength 
  is fixed to $2\pi$.} \label{fig:speed}
\end{figure}

\begin{figure}
  \centering
  \includegraphics[width=0.70\textwidth]{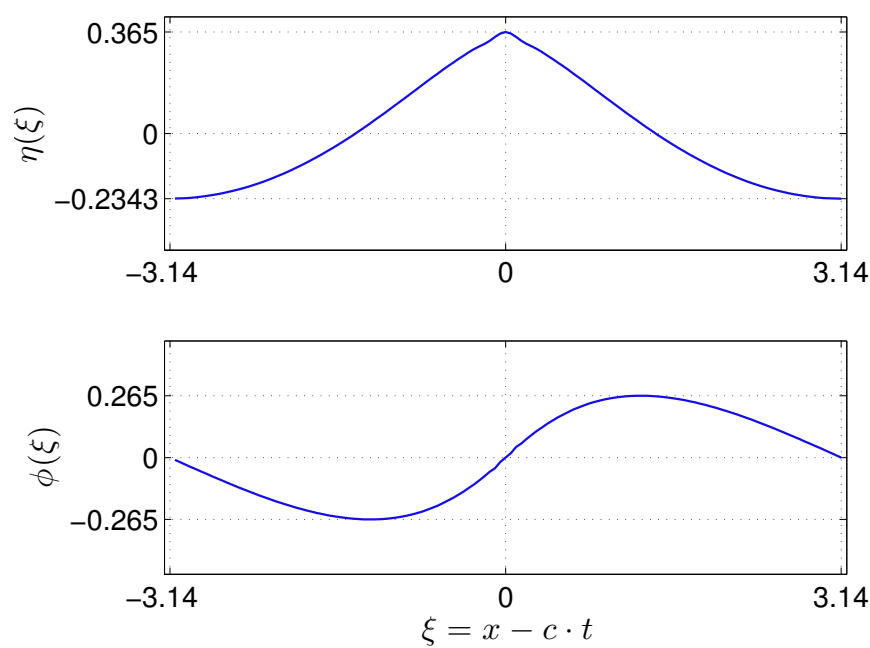}
  \caption{\small\em Periodic travelling wave to the \acs{gkg} equations for the steepness 
  parameter $\eps = 0.29967$.}
  \label{fig:limit}
\end{figure}

The comparison for various steepnesses is shown on Figure~\ref{fig:steady}. We recall that the steepness $\eps$ of a periodic wave is defined as
\begin{equation*}
  \eps\ \equiv\ \half\,(a^+ - a^-)\kappa, \qquad
  a^+\ \equiv\ \max_x\{\eta(x)\}, \qquad
  a^-\ \equiv\ \min_x\{\eta(x)\}.
\end{equation*}
One can see on Figure~\ref{fig:steady} that the differences with the reference solution \eqref{eq:stokes}--\eqref{eq:stokes3} are unnoticeable (to the graphical resolution) up to $\eps \sim 0.29$. Additionally to the shape, it is also instructive to compare the speed of travelling wave propagation with respect to the exact asymptotic result \eqref{eq:stokes3}. The comparison is shown on Figure~\ref{fig:speed}. The agreement up to very high steepnesses $\eps \sim 0.28$ is excellent. By using the numerical continuation techniques, we can compute the travelling wave solutions up to $\eps \sim 0.30$. It is interesting that at this steepness the periodic wave starts to develop an angular singularity at the crest similarly to the classical limiting Stokes wave theory \cite{Stokes1847}. This shape is represented on Figure~\ref{fig:limit}. However, for the full Euler equations the limiting wave steepness is equal to $\eps \sim 0.4432$ \cite{Williams1981}. Nevertheless, we find the agreement to be quite satisfactory considering that the \acs{gkg} model has not been designed to represent such extreme solutions.

\begin{figure}
  \centering
  \subfigure[\small\em Long perturbation]{%
\includegraphics[width=0.48\textwidth]{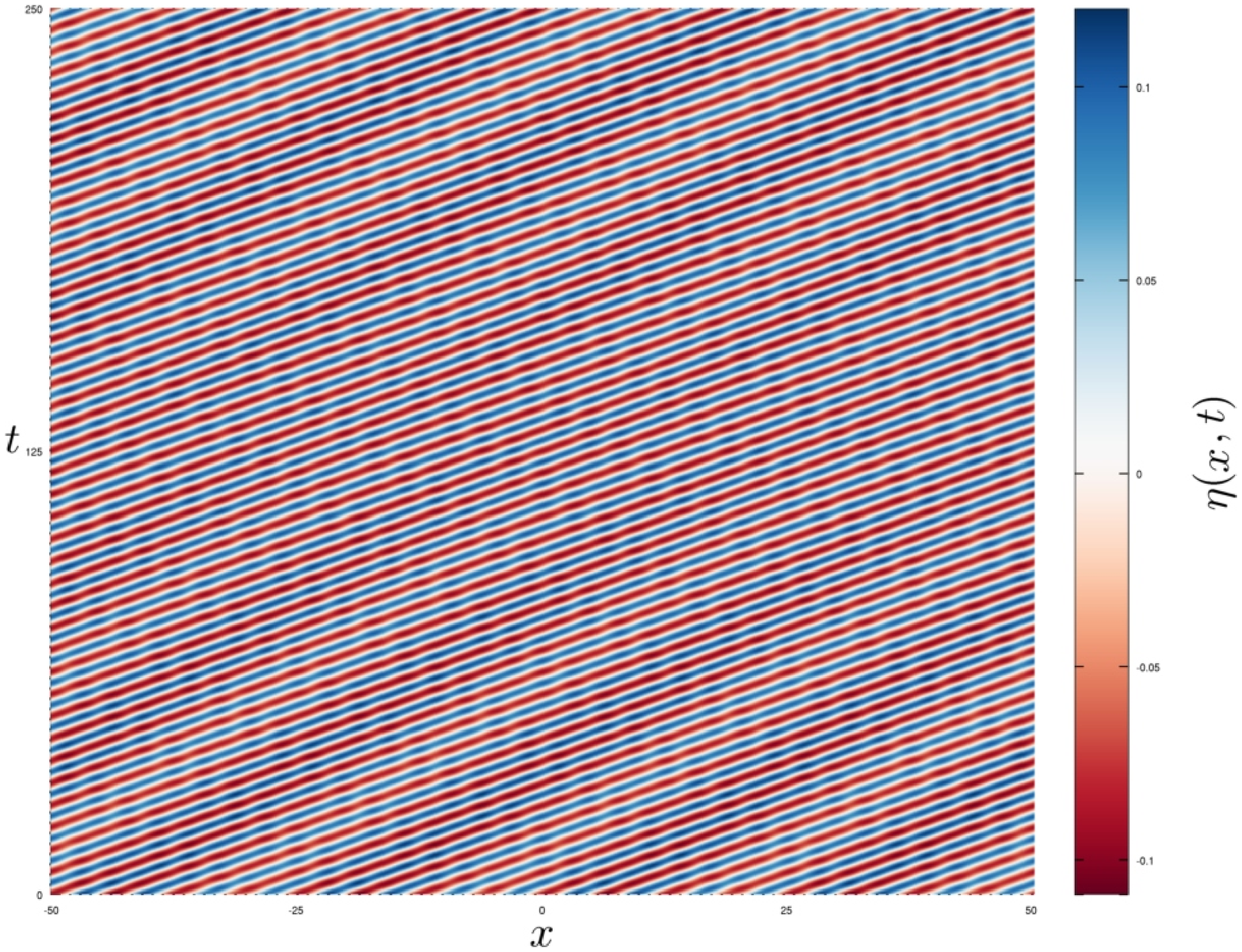}}
  \subfigure[\small\em Short perturbation]{%
\includegraphics[width=0.48\textwidth]{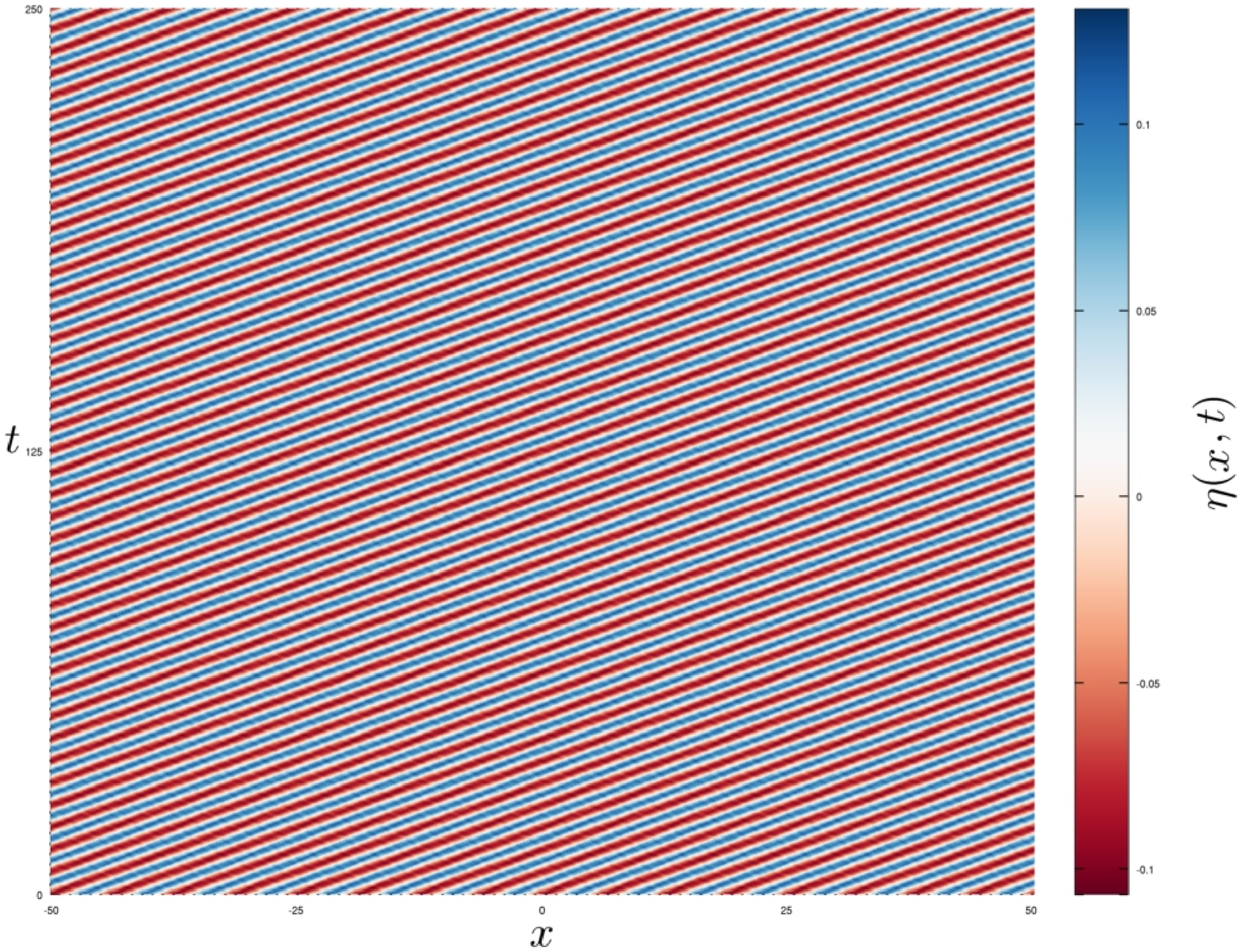}}
  \caption{\small\em Evolution of $16$ wavelengthes of computed periodic travelling waves for $\eps = 0.0954$ during about $160$ periods: (a) long wave perturbation; (b) short wave perturbation.}
  \label{fig:stokesdyn}
\end{figure}

In order to validate further the computed travelling wave profiles, we use the dynamic solver described in Section~\ref{sec:nums}. Consider the computational domain composed of $16$ periodic waves with steepness $\eps = 0.095$. The discretization was done with $N = 4096$ Fourier modes. This initial condition was propagated up to $T = 250$, which corresponds to approximatively $\sim 40$ wave periods. As one can see on Figure~\ref{fig:stokesdyn}, the initial wave system propagates uniformly in space without changing its shape. This simulation shows again that travelling waves were computed correctly. To test the stability of these solutions we consider the same initial condition with a long ($\sim 4$ wavelengths) and short ($\sim 1/4$ wavelength) wave perturbations. Both situations were simulated numerically on the same time scale and results are presented on Figures~\ref{fig:stokesdyn}(\textit{a,b}). We can see that the travelling wave solutions in the \acs{gkg} equations appear to be stable. However, a more detailed study is needed to answer this question with more certitude.

\subsection{Enveloppe soliton}

\begin{figure}
  \centering
  \includegraphics[width=0.80\textwidth]{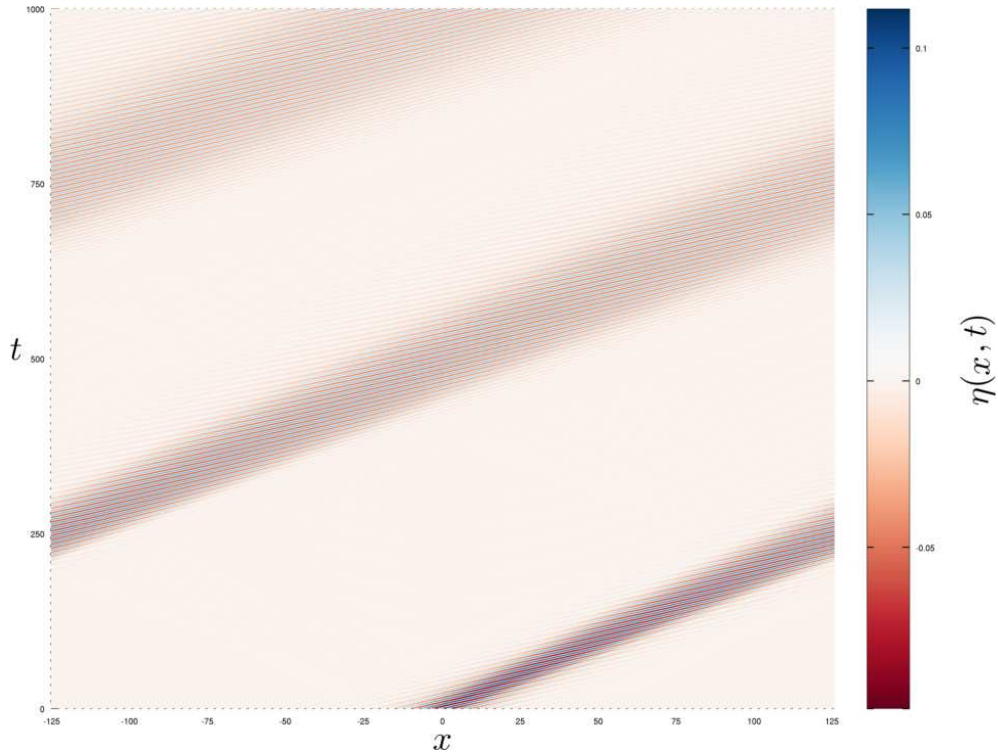}
  \caption{\small\em Space--time plot of a localized wave packet under the \acs{gkg} dynamics.}
  \label{fig:spacetime}
\end{figure}

\begin{figure}
  \centering
  \subfigure[$\sqrt{\frac{g}{\kappa}}\,t = 0.0$]{%
  \includegraphics[width=0.48\textwidth]{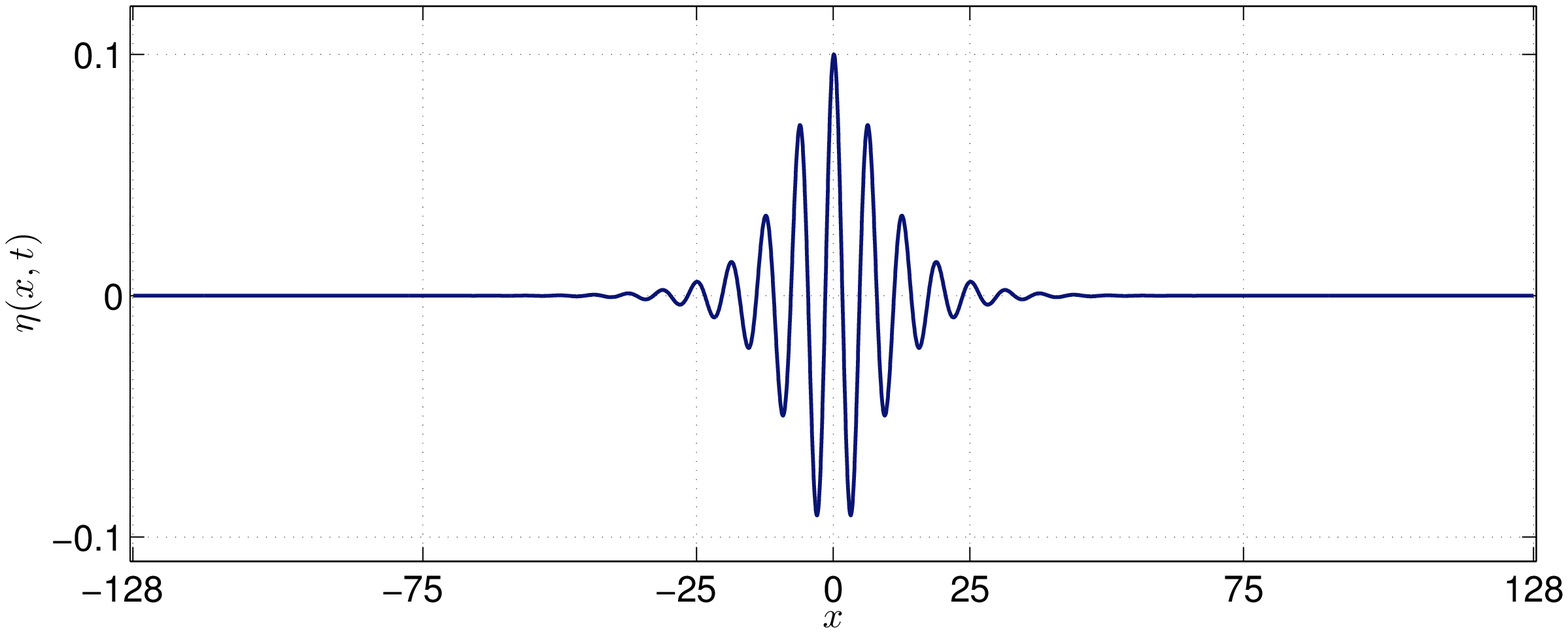}}
  \subfigure[$\sqrt{\frac{g}{\kappa}}\,t = 125.0$]{%
  \includegraphics[width=0.48\textwidth]{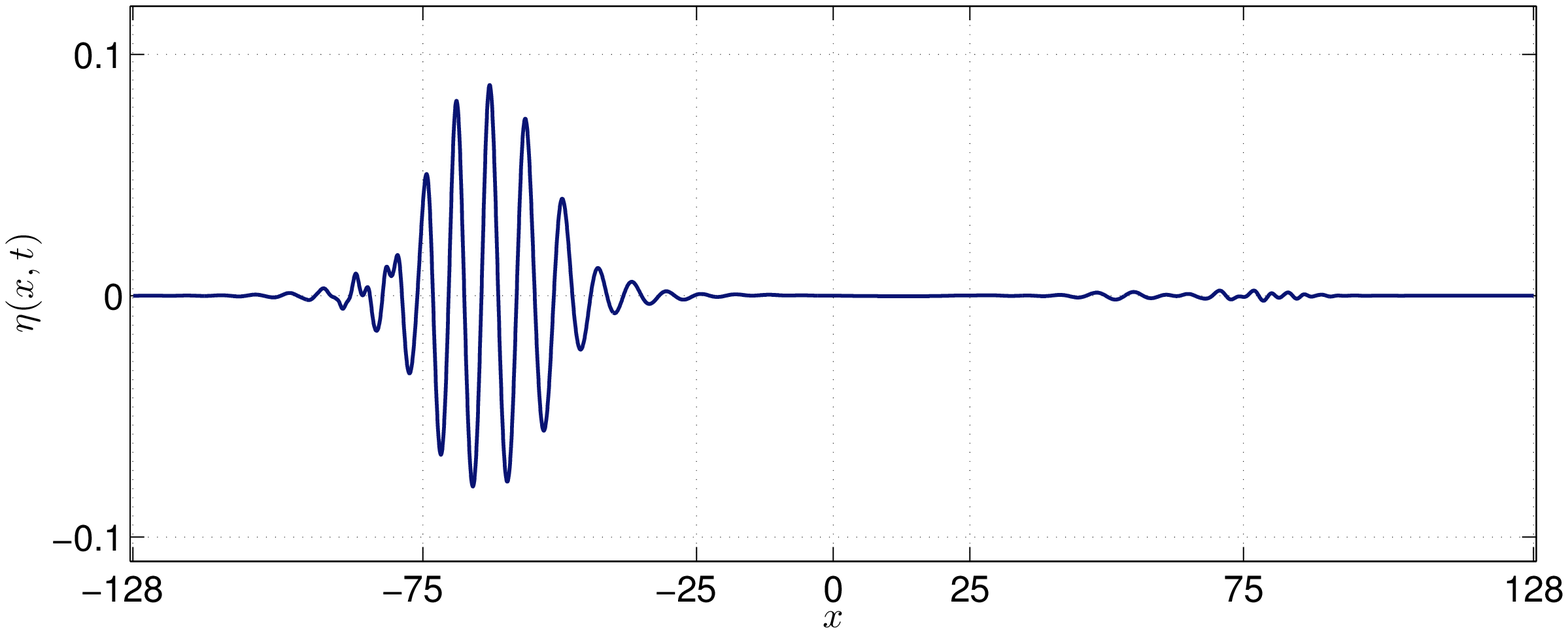}}
  \subfigure[$\sqrt{\frac{g}{\kappa}}\,t = 500.0$]{%
  \includegraphics[width=0.48\textwidth]{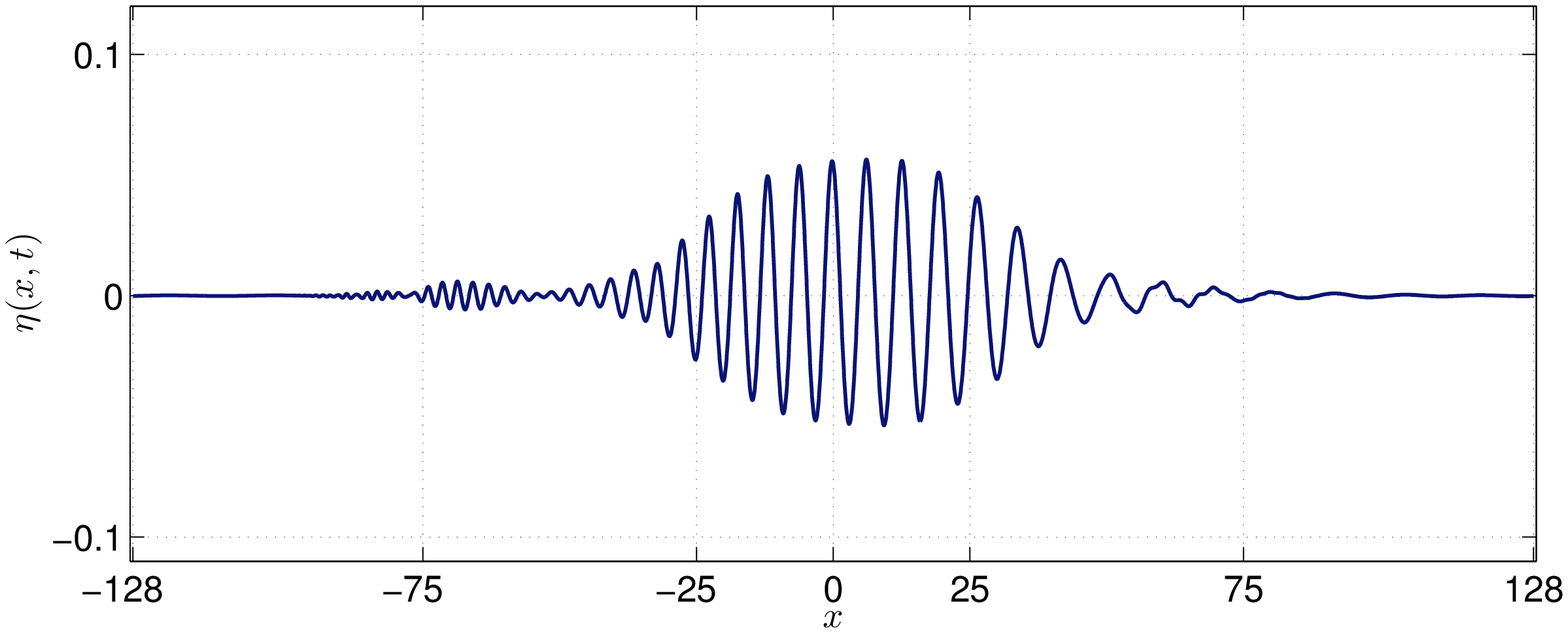}}
  \subfigure[$\sqrt{\frac{g}{\kappa}}\,t = 1000.0$]{%
  \includegraphics[width=0.48\textwidth]{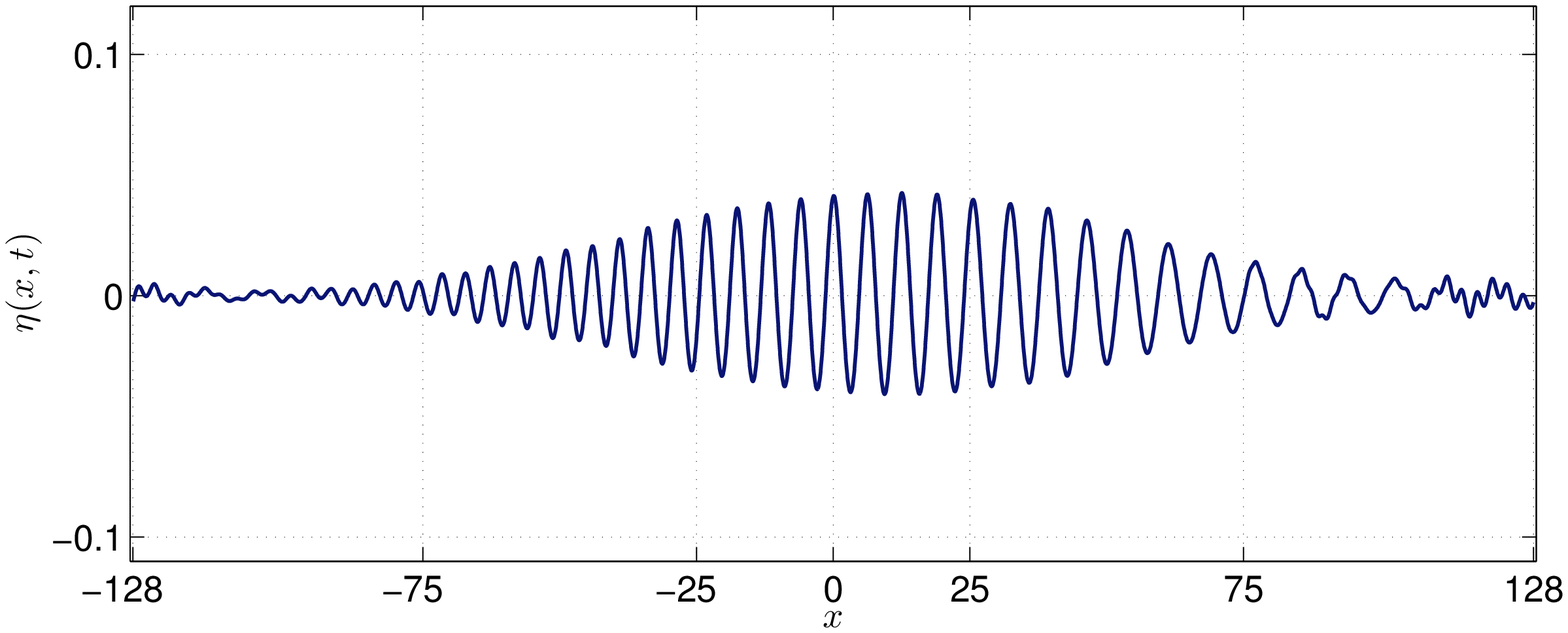}}
  \caption{\small\em Evolution of initially localized wave packet under the \acs{gkg} dynamics.}
  \label{fig:envelope}
\end{figure}

In this Section, we consider an example stemming from the wave packet propagation theory on deep waters. As it was shown for the first time by Zakharov \cite{Zakharov1968}, the free surface complex envelope $A(x,t)$ is governed by the classical \acf{NLS} \cite{Clamond2006, Henderson1999, Zakharov1968}:
\begin{equation}\label{eq:nls}
  A_t\ +\ c_g\, A_x\ +\ \frac{\ui\,c_g}{4\,k_0}\,A_{xx}\ +\ \frac{\ui\,\omega_0\,k_0^2}{2}\, A\,|A|^2\ =\ 0,
\end{equation}
where $\omega_0 = \sqrt{gk_0}$ and $c_g = \partial\omega_0/\partial k_0=\omega_0/2k_0$ is the linear group velocity. Equation \eqref{eq:nls} admits the envelope soliton solution:
\begin{equation}\label{eq:soliton}
  A(x,t)\ =\ a\,\sech\!\left(\sqrt{2}k_0^2(x-c_g t)\right)\exp(-\ui a^2 k_0^2 \omega_0t/4).
\end{equation}
The free surface elevation $\eta(x,t)$ and the velocity potential $\phi(x,t)$ can be recovered from the complex envelope $A(x,t)$ in the following way:
\begin{equation}\label{eq:env}
  \eta(x,t)\ =\ \Re\bigl\{A(x,t)\,\ue^{\ui(k_0x - \omega_0 t)}\bigr\}, \qquad
  \phi(x,t)\ =\ \Re\Bigl\{-\frac{\ui\omega_0}{k_0}A(x,t)\,\ue^{\ui(k_0x - \omega_0 t)}\Bigr\}.
\end{equation}
The evolution of this envelope soliton in higher-order models was studied in \cite{Clamond2006, Henderson1999}. Consequently, we put this localised structure as the initial condition in the \acs{gkg} equations. Consider the computational domain $[-128, 128]$ with periodic boundary conditions and the envelope soliton \eqref{eq:soliton} (transformed to physical variables using formulas \eqref{eq:env}) with $a = 0.1$, $\kappa \equiv k_0 = 1.0$ and $g = 1$. We simulated the evolution of this wave packet until $T = 1000.0$ which was sufficient for the packet to go around the computational domain three times. The space-time evolution is shown on Figure~\ref{fig:spacetime} and several individual snapshots of the free surface elevation are shown on Figure~\ref{fig:envelope}. The shape of the envelope soliton is not preserved exactly, of course. However, during short times the preservation is satisfactory. On snapshots \ref{fig:envelope} (b \& c) one can notice a small wavelet travelling in the opposite direction. The general effect is the broadening of the wave packet in agreement with previous investigations \cite{Shemer2009, Shemer2010a}.

\subsection{Shock wave formation}

\begin{figure}
  \centering
  \subfigure[$t = 0$ $\s$]%
  {\includegraphics[width=0.80\textwidth]{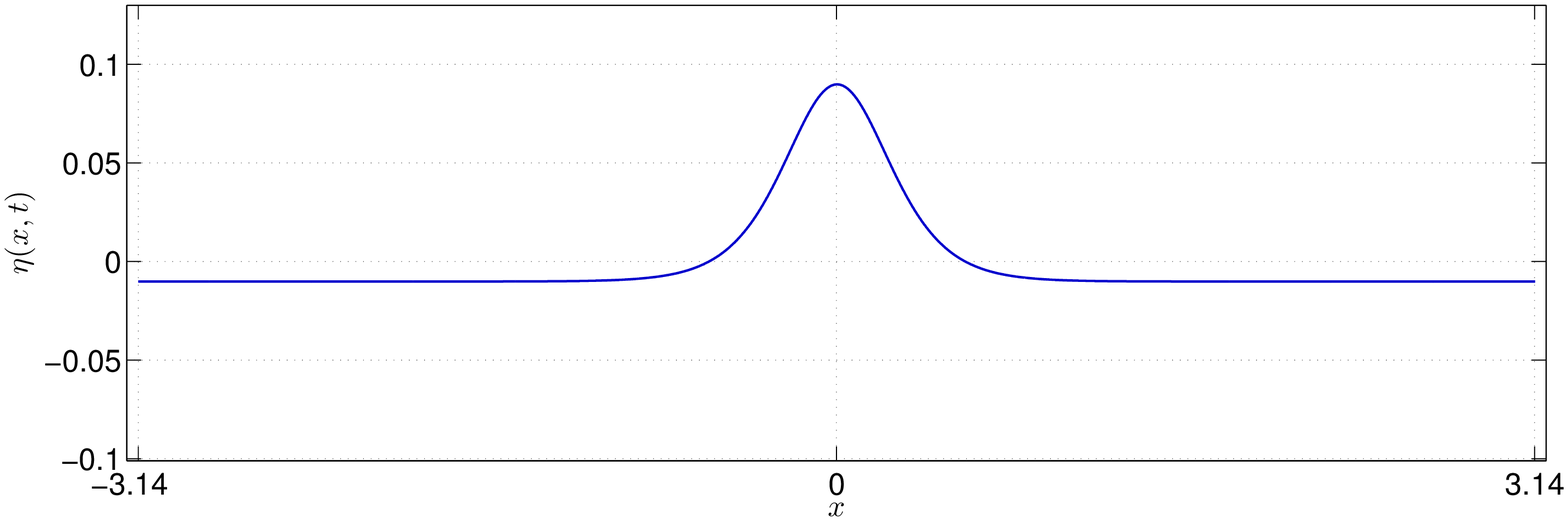}}
  \subfigure[$t = 2.875$ $\s$]%
  {\includegraphics[width=0.80\textwidth]{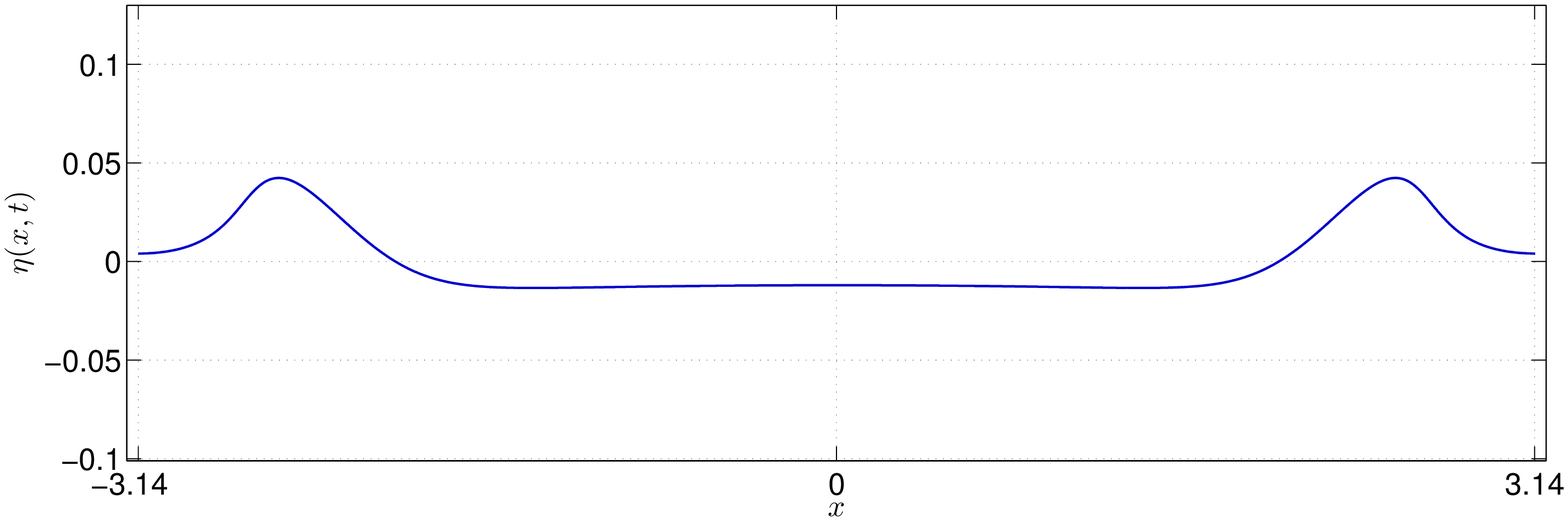}}
  \subfigure[$t = 5.75$ $\s$]%
  {\includegraphics[width=0.80\textwidth]{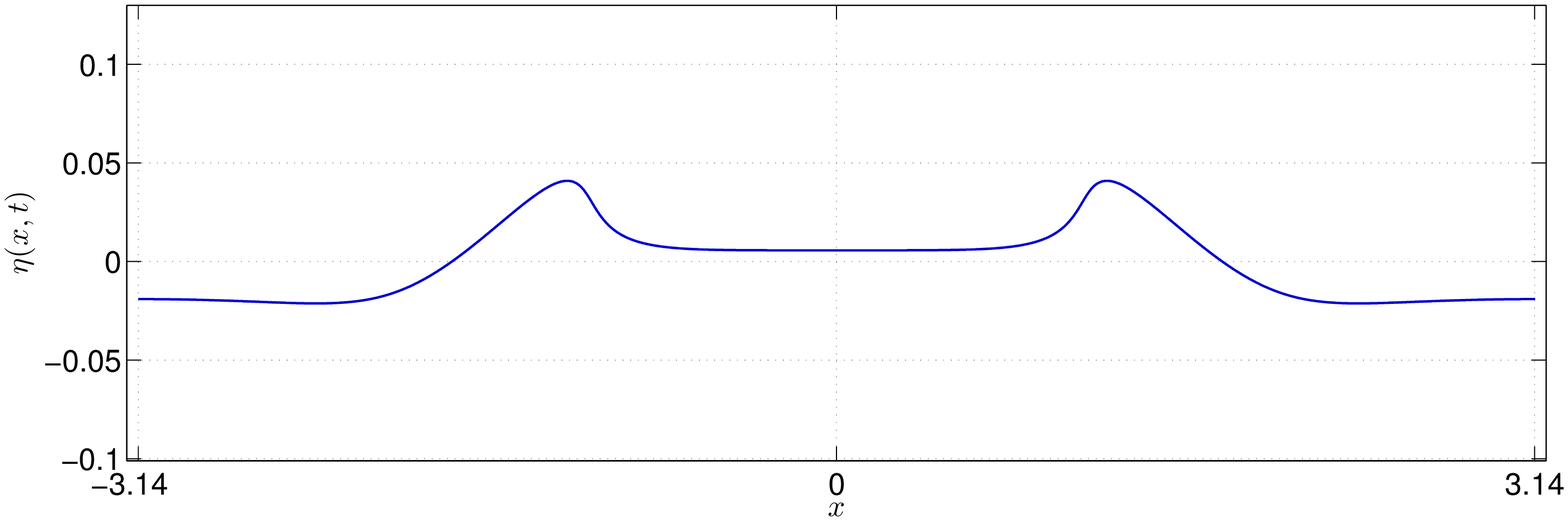}}
  \subfigure[$t = 8.625$ $\s$]%
  {\includegraphics[width=0.80\textwidth]{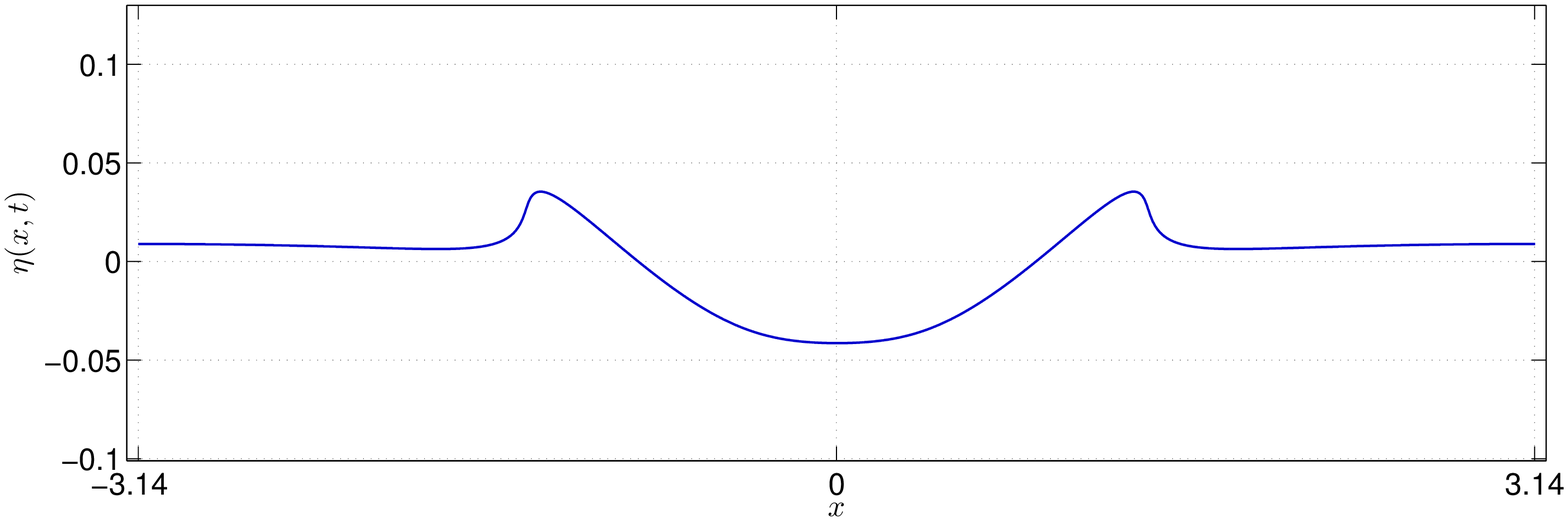}}
  \caption{\small\em Several snapshots of an initial bump evolution. See also Figure~\ref{fig:stbump}. Free surface at the final simulation time is shown on Figure~\ref{fig:shock}.}
  \label{fig:bshots}
\end{figure}

\begin{figure}
  \centering
  \includegraphics[width=0.78\textwidth]{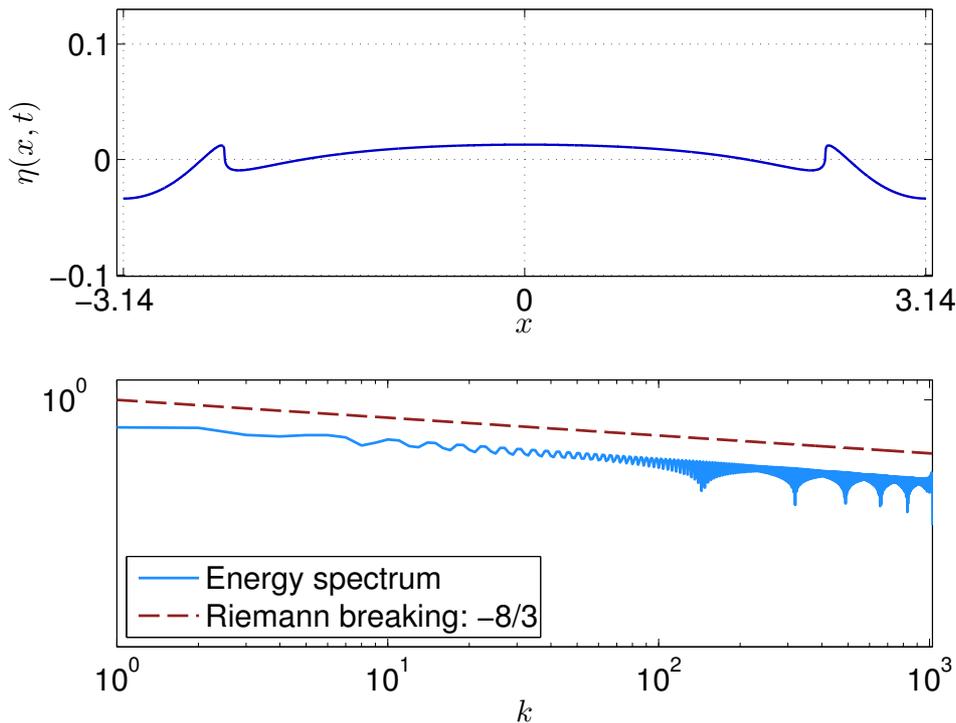}
  \caption{\small\em Free surface elevation and the energy spectrum at the final simulation time $T = 11.5$ $\s$. The red dotted line shows the theoretical prediction of a Riemann wave breaking spectrum \cite{Muraki2007, Pelinovsky2013a}.}
  \label{fig:shock}
\end{figure}

\begin{figure}
  \centering
  \includegraphics[width=0.78\textwidth]{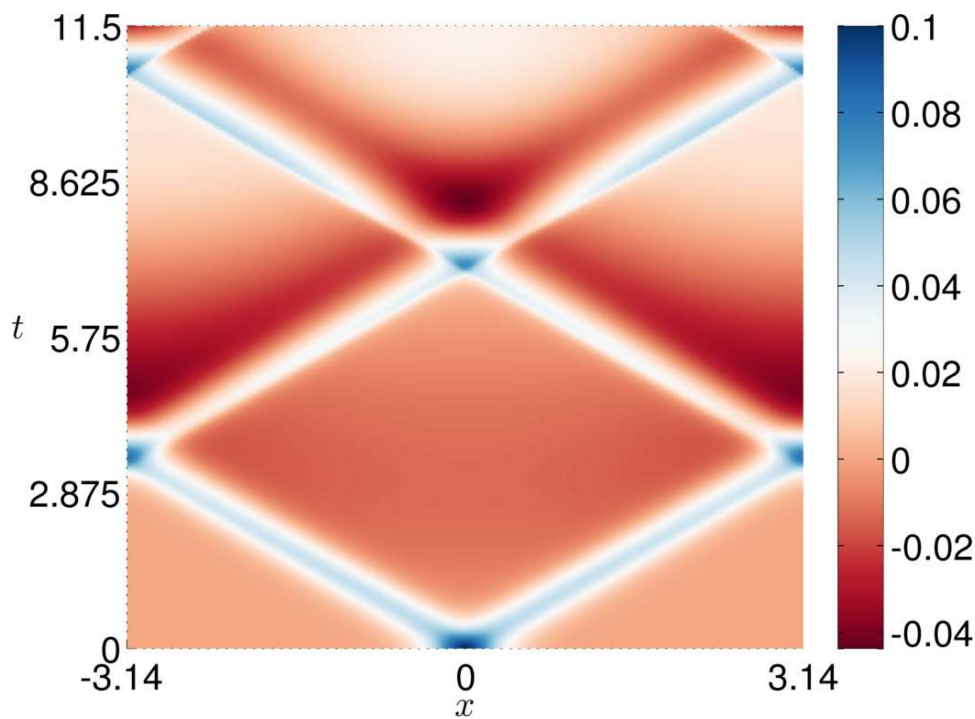}
  \caption{\small\em Space--time dynamics of an initial bump posed on the free surface in the \acs{gkg} equations.}
  \label{fig:stbump}
\end{figure}

Finally, we present an additional test-case where the \acs{gkg} system shows an interesting behaviour. The initial condition is taken to be a localised bump on the free surface with zero initial velocity:
\begin{equation*}
  \eta(x,0)\ =\ a\,\sech^2(kx), \qquad \phis(x,0)\ =\ 0.
\end{equation*}
All the values of physical and numerical parameters are given in Table~\ref{tab:breaking}. The space-time dynamics of this system is shown on Figure~\ref{fig:stbump} and several snapshots of the free surface elevation are depicted on Figure~\ref{fig:bshots}. The particularity of this simulation consists in two shock waves which develop at the free surface. The snapshot at the final simulation time $T$ is shown on the upper panel of Figure~\ref{fig:shock}. One can clearly see the sharp transitions at the free surface. It is even more instructive to look at the energy spectrum which is depicted on the bottom panel of the same Figure. For the sake of comparison, we plot also the energy spectrum of a breaking Riemann wave which was recently shown to be exactly of the form $|\hat{\eta}_k|^2 \sim k^{-8/3}$ \cite{Muraki2007, Pelinovsky2013a}. This excellent agreement shows that the \acs{gkg} system may produce wave breaking of the similar type as classical shallow-water type systems. This result was to be expected since the \acs{gkg} system is a deep water counterpart of the classical Saint-Venant equations \cite{SV1871}.

\begin{table}
  \centering
  \begingroup\setlength{\fboxsep}{0pt}
  \colorbox{lightgray}{
  \begin{tabular}{l|c}
  \hline\hline
  Gravity acceleration: $g$ [$\mathsf{m}\,\mathsf{s}^{-2}$] & $1.0$ \\
  Characteristic wavenumber: $\kappa$ [$\mathsf{m}^{-1}$] & $0.7$ \\
  Computational domain half-length: $\ell$ [$\mathsf{m}$] & $\pi$ \\
  Final simulation time: $T$ [$\s$] & $11.5$ \\
  Initial condition amplitude: $a$ [$\mathsf{m}$] & $0.1$ \\
  Initial bump width: $k$ [$\mathsf{m}^{-1}$] & $\pi$ \\
  Number of Fourier modes: $N$ & 4096 \\
  \hline\hline
  \end{tabular}}\endgroup
  \caption{\small\em Physical and numerical parameters used for the simulation of the shock wave formation in \acs{gkg} equations.}
  \label{tab:breaking}
\end{table}

\section{Conclusions and future work}\label{sec:concl}

We discussed the derivation of some \acf{gkg} equations, which are a new model for water waves propagating in deep water approximation. This model already appeared as an illustration for the relaxed variational formulation \cite{Clamond2009}. However, in the present study, the structure of this model is further investigated and a multi-symplectic formulation was proposed. Moreover, we computed periodic travelling wave solutions and we showed that they approximate fairly well the corresponding solutions of the full Euler equations, including the formation of a limiting wave with a singular point at the crest \cite{Maklakov2002, Stokes1847}.

The dynamics of regular periodic waves was studied and these solutions appear to be stable under long and short wave perturbations. Finally, we showed also that solutions of \acs{gkg} equations may produce the shock wave formation phenomenon of the similar type as the breaking of Riemann waves in shallow water models \cite{Muraki2007, Pelinovsky2013a}. To our knowledge, it is the first approximate model in deep waters which shows this behaviour. 

Although the results presented in this paper are encouraging, further investigations would be necessary to assess the relevance and limitations of the \acf{gkg} for modelling water waves in deep water. The comparisons with the compact Dyachenko--Zakharov equation \cite{Dyachenko2011, Fedele2012a} might be interesting. However, the \acs{gkg} equations may also be a relevant model in contexts different from water waves.

Concerning the perspectives, the stability of periodic travelling wave solutions to the \acs{gkg} equations has to be properly studied using the Floquet theory \cite{Deconinck2006}. Moreover, all the simulations presented hereinabove were performed in (1+1)D wave propagation. Similar numerical 
analysis has to be performed in (2+1)D as well.

\subsection*{Acknowledgments}
\addcontentsline{toc}{section}{Acknowledgments}

D.~\textsc{Dutykh} would like to acknowledge the support from ERC under the research project ERC-2011-AdG 290562-MULTIWAVE and the hospitality of the Laboratory J.~A.~Dieudonn\'e, University of Nice -- Sophia Antipolis during his visits.

\addcontentsline{toc}{section}{References}
\bibliographystyle{abbrv}
\bibliography{biblio}

\end{document}